\pdfoutput=1

\documentclass[11pt]{article}
\usepackage{comment}
\usepackage{verbatim}
\usepackage{bbm}
\usepackage{tikz}
\usepackage{multirow, makecell}
\usepackage{amsmath}
\usepackage{graphicx} 
\usepackage{wrapfig}
\usepackage{subcaption}
\usepackage{xcolor,colortbl}
\usepackage{algorithm}
\usepackage{algorithmic}

\definecolor{Gray}{gray}{0.9}
\newcolumntype{a}{>{\columncolor{Gray}}c}

\usepackage[utf8]{inputenc} 
\usepackage[T1]{fontenc}    
\usepackage{hyperref}       
\usepackage{url}            
\usepackage{booktabs}       
\usepackage{amsfonts}       
\usepackage{nicefrac}       
\usepackage{microtype}      

\usepackage[preprint]{acl}

\usepackage{times}
\usepackage{latexsym}

\usepackage[T1]{fontenc}

\usepackage[utf8]{inputenc}

\usepackage{microtype}

\usepackage{inconsolata}

\usepackage{graphicx}

\title{Corpus Poisoning via  Approximate Greedy Gradient Descent}
\author{Jinyan Su$^1$, 
  Preslav Nakov$^2$,  Claire Cardie$^1$\\
  $^1$Department of Computer Science, Cornell University\\
  $^2$Mohamed bin Zayed University of Artificial Intelligence\\
  \texttt{\{js3673,ctc9\}@cornell.edu, preslav.nakov@mbzuai.ac.ae}
  }

\begin{document}
\maketitle
\begin{abstract}
Dense retrievers are widely used in information retrieval and have also been successfully extended to other knowledge intensive areas such as language models, e.g., Retrieval-Augmented Generation (RAG) systems.  Unfortunately, they have recently been shown to be vulnerable to corpus poisoning attacks
in which a malicious user  injects a small fraction of adversarial passages into the retrieval corpus to trick the system into returning these passages among the top-ranked results for a broad set of user queries. Further study is needed to understand the extent to which these attacks could limit the deployment of dense retrievers in real-world applications.
In this work, we propose Approximate Greedy Gradient
Descent (AGGD), a new attack on dense retrieval systems  based on the widely used HotFlip method for efficiently generating adversarial passages.  We demonstrate  that AGGD can select a higher quality set of token-level perturbations than  HotFlip by replacing its random token sampling with a more structured search. Experimentally, we show that our method achieves a high attack success rate on several datasets and using several retrievers, and can generalize to unseen queries and new domains. Notably, our method is extremely effective in attacking the ANCE retrieval model, achieving attack success rates that are 15.24\% and 17.44\% higher on the NQ and MS MARCO datasets, respectively, compared to HotFlip. Additionally, we demonstrate AGGD's potential to replace HotFlip in other adversarial attacks, such as knowledge poisoning of RAG systems.
\end{abstract}

\section{Introduction}
Dense retrievers, despite their wide application and extensive deployment in real-world systems \citep{wan2022fast, mitra2017learning, lewis2020retrieval, guu2020retrieval, qu2020rocketqa}, have recently been shown to be vulnerable to various adversarial attacks such as corpus poisoning attacks \citep{zhong2023poisoning} and  data poison attacks \citep{long2024backdoor, liu2023black}, raising concerns about their security.
Given that the corpora used in retrieval systems are often sourced from openly accessible platforms like Wikipedia and Reddit, a concerning scenario arises in which malicious actors can poison the retrieval corpus by injecting some adversarial passages, fooling the system into retrieving these malicious documents rather than the most relevant ones. Such attacks might be used for search engine optimization \citep{patil2013search} for promoting advertisement, or disseminating disinformation and hate speech.

A conventional approach for such attacks is HotFlip \citep{ebrahimi2017hotflip}, which involves collecting a candidate set for a single randomly sampled token position and finding the best token in the candidate set with which to replace. In addition to corpus poisoning attacks on dense retrieval systems, HotFlip has been widely used in many other settings, such as knowledge poisoning attacks on retrieval augmented generation (RAG) systems \citep{zou2024poisonedrag} and adversarial prompt generation \citep{zou2023universal}.

In this work, we begin by thoroughly investigating the HotFlip attack on dense retrieval systems to identify its limitations. Based on these insights, we propose a new general attack method called Approximate Greedy Gradient Descent (AGGD). Our experimental results show that AGGD can perform corpus poisoning attacks on dense retrieval systems more effectively, revealing their vulnerability. Though we use corpus attacks on dense retrievers as our primary example, it is important to note that AGGD can replace Hot Flip as a whole in any attack scenarios where HotFlip is applicable.

The main difference between AGGD and Hot Flip is that AGGD uses gradient information more effectively by selecting the top-ranked token from all token positions, rather than over a single randomly sampled position. This approach makes AGGD's search trajectory deterministic, enabling a more structured best-first search.
Experimental results demonstrate that AGGD achieves a high attack success rate across various datasets and retrieval models.
In summary, our contributions are
\begin{itemize}
\item We provide a thorough understanding of the existing HotFlip adversarial attack method, explaining its mechanics and identifying its potential problems.
\item We propose AGGD, a gradient-based method that replaces a randomized greedy search with a systematic best-first greedy search over the discrete token space. We demonstrate the effectiveness of AGGD in various settings. 

\item We conduct extensive experiments to show the vulnerability of dense retrievers under AGGD. For example, when attacking the ANCE retriever, injecting just one adversarial passage  can achieve an attack success rate of 80.92\% and 65.68\% on these datasets, respectively, improving by 15.24\% and 17.44\% over HotFlip. The generated adversarial passage also possesses the capability to transfer to unseen queries in other domains.

\end{itemize}

\section{Related Work}
\textbf{Dense Retrieval}
Dense retrievers utilize dense vector representations to capture the semantic information of passages and have demonstrated tremendous effectiveness compared to traditional retrieval systems \citep{yates2021pretrained}. Consequently, they have been employed in many knowledge-intensive areas such as information retrieval \citep{karpukhin2020dense, gillick2019learning, wu2019scalable, wan2022fast, mitra2017learning}, open-domain question answering and language model pre-training \citep{lewis2020retrieval, guu2020retrieval, qu2020rocketqa}. 
For instance, retrieval-augmented generation (RAG) models
\citep{lewis2020retrieval, guu2020retrieval, lee2019latent} combine language models with a retriever component to generate more diverse, factual and specific content.

 \textbf{Adversarial Attacks in Retrieval Systems}
Black-hat search engine optimization, which aims to increase the exposure of certain documents through malicious manipulation, poses a threat by reducing the quality of search results and inundating users with irrelevant pages \citep{castillo2011adversarial, liu2023black}. Previous work has shown that retrieval systems  are susceptible to small perturbations:  making small edits to a target passage can significantly alter its retrieval rank \citep{song-etal-2020-adversarial, raval2020one, song2022trattack} for individual or a small set of queries. More recently, a stronger setting known as \textit{corpus poisoning} attack has been studied in \cite{zhong2023poisoning}, where the attack success rate of an adversarial passage is evaluated on unseen queries rather than on targeted given queries. These attacks differ from data poisoning attacks \citep{long2024backdoor, liu2023black, chen2017targeted, schuster2020humpty}, as adversarial passages are injected into the retrieval corpus rather than the training data of the retrievers. The retrieval model remains unchanged in a corpus poisoning attack. 

 \textbf{Discrete Optimization}
Many adversarial attacks in NLP involve discrete optimization, whether in classification tasks \citep{wallace2019universal, ebrahimi2017hotflip, song2021universal}, retrieval systems \citep{jia2017adversarial, song-etal-2020-adversarial, raval2020one, song2022trattack}, or adversarial prompt generation \citep{zou2023universal, shin2020autoprompt, wen2024hard}. 
The objective is to find small perturbations to the input to lead the model to make erroneous predictions. 
Due to the discrete nature of texts, directly applying adversarial attack methods from prior computer vision research \citep{xiao2021you, tolias2019targeted} is infeasible. Instead, many methods build upon HotFlip \citep{ebrahimi2017hotflip} and approximate the effect of replacing a token using gradients. 
 



\section{Motivation}

In this section we motivate our approach by formalizing the corpus poisoning problem setting (Section~\ref{subsec:corpus-poisoning}), describing and analyzing the standard HotFlip approach for producing adversarial passages \citep{zhong2023poisoning} (Section~\ref{subsec:hot-flip}), and identifying a potential problem with this approach (Section~\ref{subsec:problem}).

\subsection{Corpus Poisoning Problem Setting}
\label{subsec:corpus-poisoning}
In retrieval systems, the retrieval model takes a user query $q$ and returns a ranked list of the $k$ most relevant passages from a large corpus collection $\mathcal{C} = \{p_1, \cdots, p_{|\mathcal{C}|}\}$ consisting of $|\mathcal{C}|$ passages. Compared to sparse retrieval models, which rely on lexical matching, dense retrievers rely on semantic matching. Specifically, the queries and the passages are first represented by $d$-dimensional dense vectors using a query encoder $E_{q}(\cdot)$ and passage encoder $E_{p}(\cdot)$, respectively. Relevance scores can then be computed according to a similarity function. A commonly used similarity function is the dot product of the dense vector representations of the query $q$ and the passage $p$:  $\text{Sim}(q, p) = E_q(q)^T E_p(p)$.
Finally, a ranked list of the $k$ most relevant passages $L=[\tilde{d}_1, \tilde{d}_2, \cdots, \tilde{d}_k], (L\subseteq  \mathcal{C})$ is returned according to the relevance score. 

We consider the problem of corpus attacks on a dense retrieval system, where we design an algorithm to find
a small set of adversarial passages $\mathcal{A}=\{a_1, \cdots, a_{|\mathcal{A}|}\}$ that can be retrieved by as many queries as possible for query distribution $\mathcal{P}_{\mathcal{Q}}$. These adversarial passages are then inserted into the corpus $\mathcal{C}$ to fool the dense retrieval models into retrieving passages from $\mathcal{A}$ rather than the most semantically relevant passages from the original corpus $|\mathcal{C}|$. The adversarial passage set $\mathcal{A}$ should be much smaller than the original corpus $\mathcal{C}$. The attack quality of the adversarial passage set $\mathcal{A}$ is typically evaluated based on its \textit{attack success rate}, i.e., the percentage of queries for which at least one adversarial passage appears in the top-$k$ retrieval results.  

Formally, the overall objective is to find an adversarial passage $a$, that maximizes the expected similarity to a query $q$ sampled from distribution $\mathcal{P}_{\mathcal{Q}}$, i.e., 
 \begin{equation*}
 a =\arg\max_{a} \mathbb{E}_{q\sim \mathcal{P}_{\mathcal{Q}}}\text{Sim}(q, a)
 \end{equation*}
In practice, we estimate the query distribution using a training set of queries $\mathcal{Q}=\{q_1. \cdots, q_{|\mathcal{Q}|}\}$, and we aim to find an $a$ with  maximal similarity to $\mathcal{Q}$, i.e., 
 \begin{equation}\label{eq: 1}
 \small a =\arg\max_{a} \frac{1}{|\mathcal{Q}|} \sum_{q_i \in \mathcal{Q}}\text{Sim}(q_i, a)=\arg\min_{a}\ell(a)
 \end{equation}
 where $\ell(a)=-\frac{1}{|\mathcal{Q}|}\sum_{q_i\in \mathcal{Q}}\text{Sim}(q_i, a)$.
The problem setting is realistic in search engines where a malicious user might perform search  engine optimization to promote misinformation or spread spam. 

Finding the exact solution to the optimization problem (\ref{eq: 1}) is challenging  since we are optimizing over a discrete set of inputs (i.e., the tokens in a passage). Additionally, running gradient descent on the embedding space might yield solutions that exist only in the embedding space and deviate significantly from valid texts in the discrete token space. In practice, a straightforward approach that leverages the gradient w.r.t.\ the one-hot token indicators can be employed to identify a set of promising candidates for replacement \citep{zou2023universal, ebrahimi2017hotflip, shin2020autoprompt, zhong2023poisoning}. Specifically, we can compute the linearized approximation of replacing the $i$-th token $t_i$ in a passage $a$, by evaluating the gradient $\nabla_{e_{t_i}}\ell(a)$, where $e_{t_i}$ denotes the embedding of the token $t_i$. (Recall that sentence embeddings can be written as function of token embeddings, allowing us to compute the gradient with respect to the token embedding.) This idea has been adopted in many gradient-based search algorithms such as HotFlip \citep{ebrahimi2017hotflip} for producing adversarial texts,  AutoPrompt \citep{shin2020autoprompt} and Greedy Coordinate Gradient (GCG) \citep{zou2023universal} for generating prompts. We revisit the idea of HotFlip in the context of corpus attacks as an example.

\subsection{HotFlip Revisited}
\label{subsec:hot-flip}
In many text-based adversarial attacks, the goal is to find a perturbation of an input sequence --- a randomly selected passage $a$ --- to optimize some objective function. Due to the discrete nature of text, this is essentially a combinatorial search problem: the attacker searches over a class of perturbations on $a$ such as word swapping or character substitution
\citep{morris2020textattack} to produce the final adversarial text.
The corpus poisoning approach of \cite{zhong2023poisoning}, for example, aims to fool a dense retrieval system to return adversarial passages for a broad set of user queries by transforming corpus passages (i.e., sequences of tokens) into adversarial passages $a$ that exhibit maximal similarity to a query set associated with the corpus.  

Unfortunately, even if we fix the token length of a passage $|a|$ to be $m$, for a language model with vocabulary size $|\mathcal{V}|$, the total size of the search space is $|\mathcal{V}|^m$. For instance, if $m=30$, and $|\mathcal{V}|=32,500$, then the total number of sequences to evaluate is about $10^{135}$, which is computationally infeasible.

Thus, we need a method to identify a subset of the most promising perturbation sequences.
The  corpus poisoning attack  by \cite{zhong2023poisoning} 
addresses this problem using the HotFlip approach in a greedy search over candidate
token perturbations.
HotFlip \citep{ebrahimi2017hotflip}  first randomly sample a token position, and then
determines a candidate set of promising token replacements based on the gradient of their respective
one-hot input vector\footnote{The embedded version of the token is used.} representations. 
Each iteration of search involves: (1) selecting a random position in $a$; (2)
use gradient information to determine the top-$k$ token perturbation candidates; and (3) 
applying the perturbation (a token swap) that increases the similarity (decreases the loss) the most.\footnote{If no candidate increases the loss, no
change to $a$ is made.}. This process continues for a fixed number of iterations. 
\subsection{Drawbacks of HotFlip}
\label{subsec:problem}
\begin{figure}[h]
    \centering
    \includegraphics[width=0.55\textwidth]{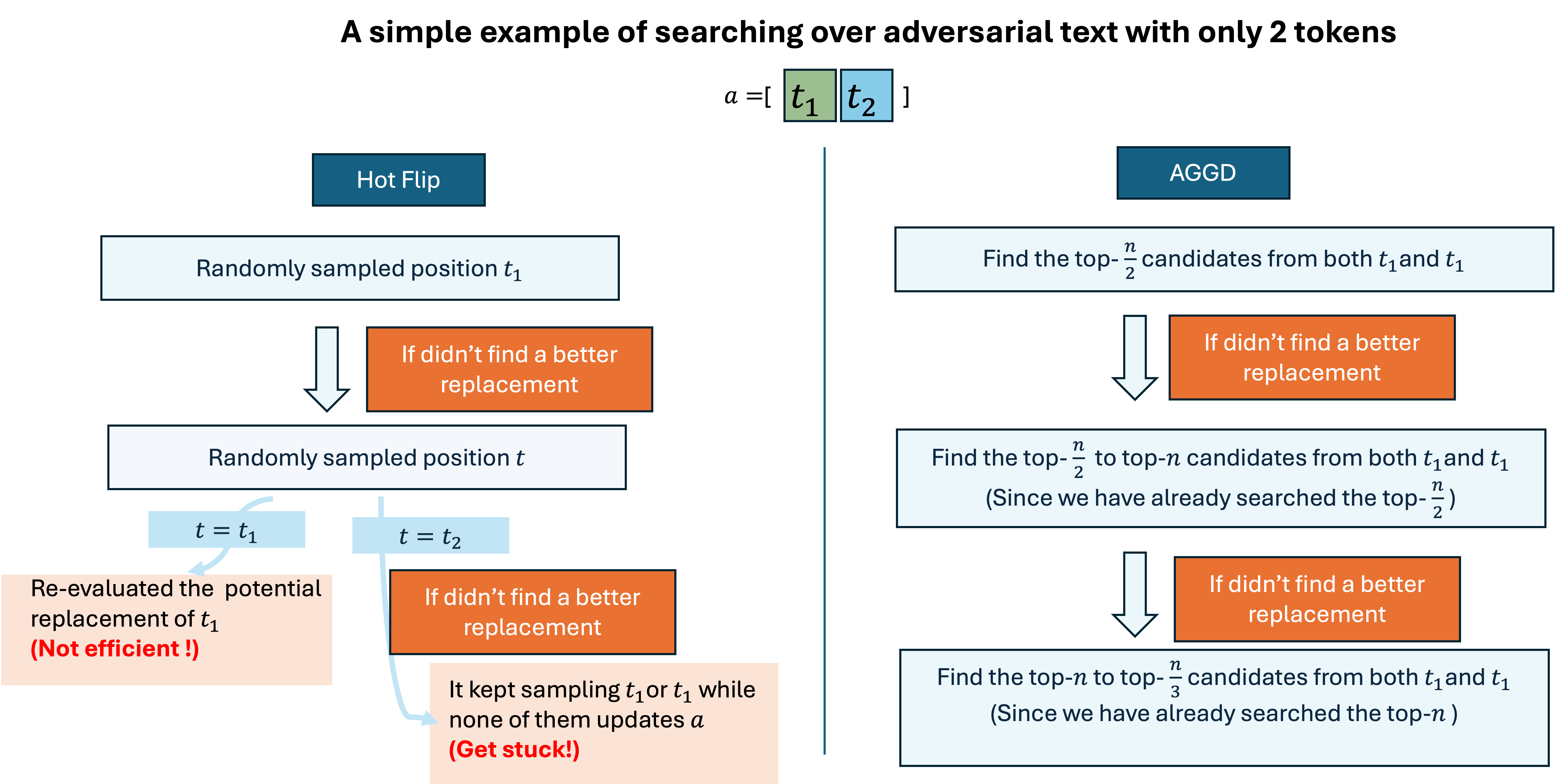}
    \caption{A simple example of finding a 2-token sequence $a$ through HotFlip (left) and AGGD (right). If HotFlip can't find a better replacement for the currently sampled token position, there is a $\frac{1}{2}$ probability that it will sample the same token position again and redo the same evaluation, which is inefficient. Moreover, if the potential replacements for another token also don't contain a better option, HotFlip continues to loop through the same search without reducing the loss. }
    \label{fig: 2-token example}
    \vspace{-0.5cm}
\end{figure}

Although HotFlip usually works well in practice, there are some noticeable problems with this approach.  For instance, it is less efficient due to its reliance on randomness for the search. Moreover, it is possible to try all token positions without finding any updates, causing HotFlip to get stuck and repeatedly search the same perturbation candidates. For example, consider an adversarial text with only two tokens  (see Figure \ref{fig: 2-token example}). If we first sample token position $t_1$ and observe no improvement for all candidates, there is still a $1/2$ probability of sampling the \textbf{same} token $t_1$
  in the next iteration. Furthermore, if both token $t_1$ and token $t_2$ have been searched and show no improvement, the process would be stuck in a loop. 
\begin{figure}[h]
    \centering
    \includegraphics[width=0.3\textwidth]{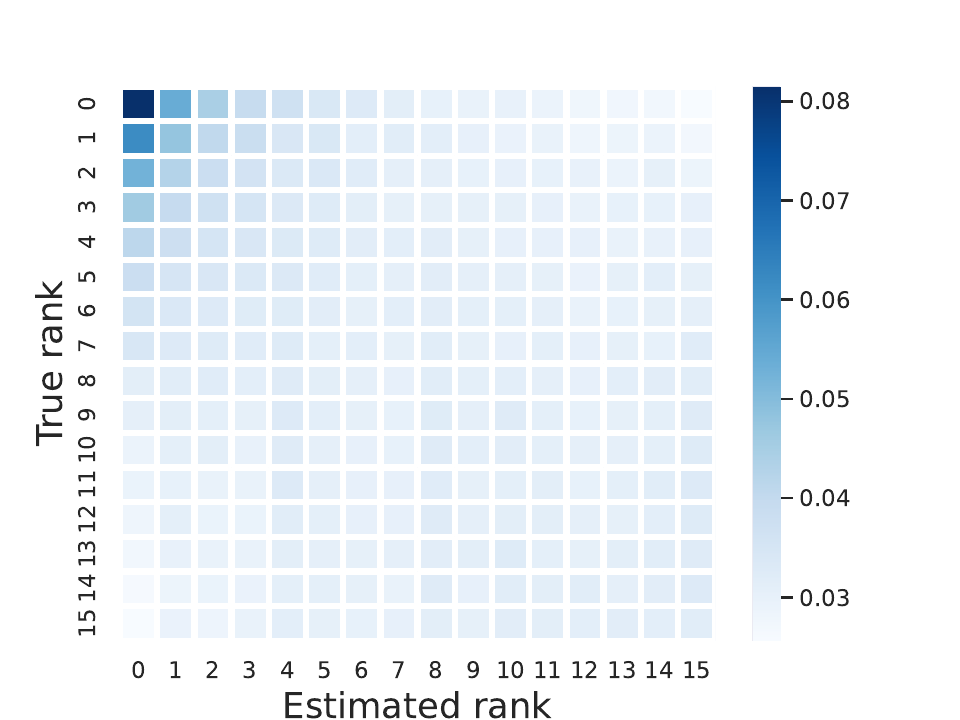}
    \caption{Comparing the true rank of words swapped with their rank according to the gradient-based Taylor approximation. The gradient identifies the top-1 correct token to swap 9\% of the time, and guesses within the top ten tokens 58\% of the time.}
    \label{fig: gradient matters}
    \vspace{-0.5cm}
\end{figure}

Motivated by these existing problems, we propose a new algorithm called Approximate Greedy Gradient Descent (AGGD)  that uses a deterministic greedy search that  
makes better use of gradient information by utilizing lower-ranked tokens(i.e., the overall most promising token swap candidates) to improve the quality of the candidate set. As shown in Figure \ref{fig: gradient matters}, most high quality potential candidates are concentrated in the low-rank area. Suppose we aim to find a text with a token length $m=30$ by maintaining a candidate set of size $n=150$. Selecting as the candidate set  the top-$5$ ranked gradients \textbf{across all token positions} is likely to result in better quality than selecting the top-$150$ ranked gradient candidates for only one token position. In the next section, we formally introduce our algorithm Approximate Greedy Gradient Descent (AGGD).




\section{Approximate Greedy Gradient Descent}
\begin{figure}[h]
    \centering    \includegraphics[width=0.5\textwidth]{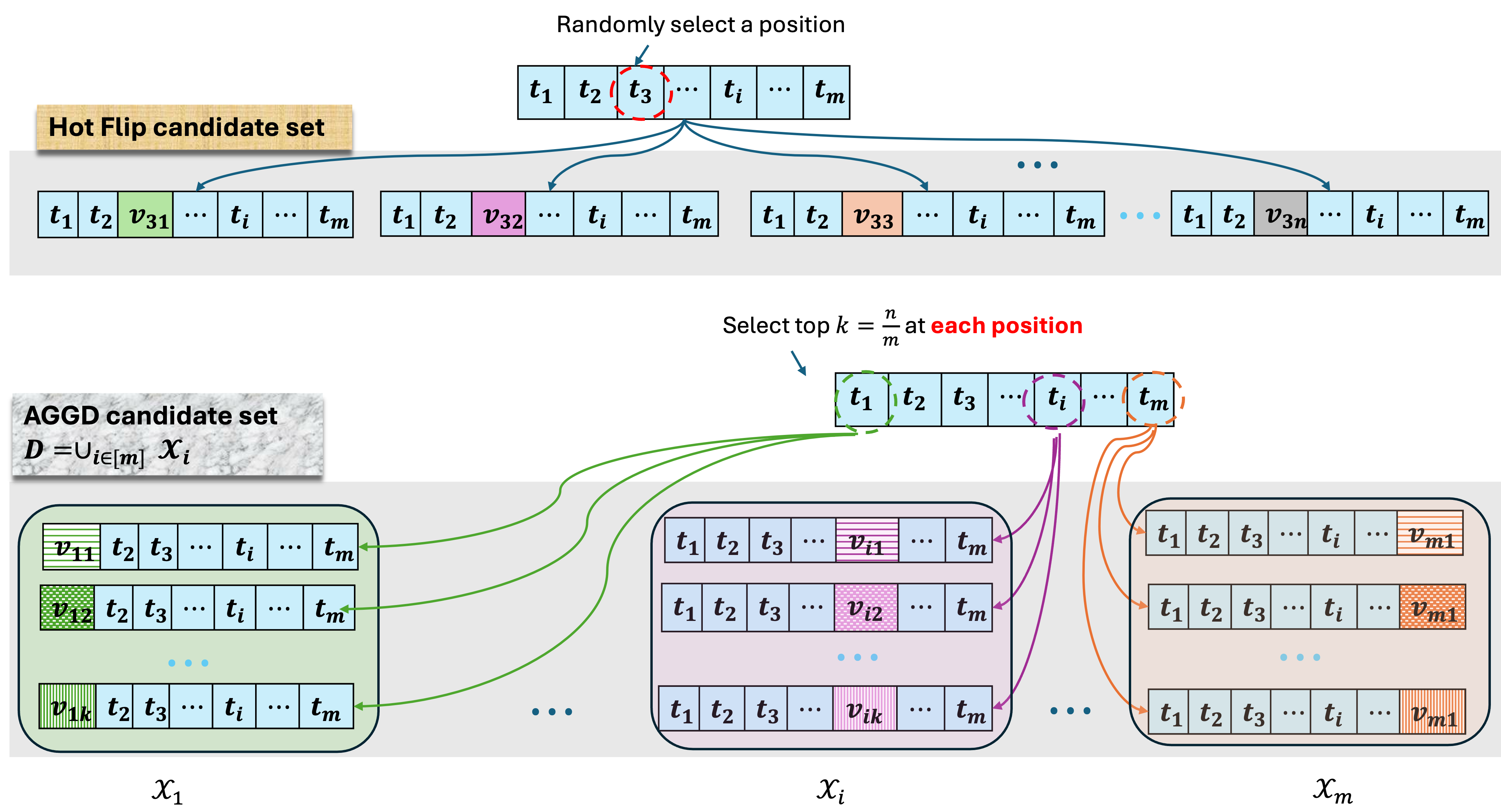}
    \caption{Illustration of HotFlip (top) and AGGD (bottom) and their candidate sets.}
    \label{fig: illustration}
    \vspace{-0.5cm}
\end{figure}

Similar to other  gradient-based search algorithms, we fist initialize the adversarial passage to be $a = [t_{1}, \cdots, t_{m}]$. We then iteratively update $a$ based on the best candidate that maximizes the similarity over batches of queries. Formally, at each step, we compute a first-order approximation of the change in the loss when swapping the $i$th token in $a$ with another token $v\in \mathcal{V}$. In contrast to Hotflip, AGGD identifies a candidate set for each of the $m$ tokens in $a$: for each token $i$, the candidate set contains the top $k=\frac{n}{m}$ ranked tokens from $\mathcal{V}$ according to the scoring function $s(v) = -e_v^T\nabla \ell(a)$, i.e., $\mathcal{X}_i = \text{top-}k_{v\in \mathcal{V}}[-e_v^T \nabla \ell(a)]$, where $e_v$ is the token embedding and the gradient is taken over the the embedding of the current adversarial passage $a$. Combining all the $\mathcal{X}_i$ leads to our overall candidate set $D^{(j)}=\cup_{i\in [m]}\mathcal{X}_i$ (as illustrated in Figure \ref{fig: illustration}). For each candidate in the set $D$, the loss is re-evaluated and $a$ is updated to the candidate with lowest loss for the next step. This requires $n$ passes of the model, which constitutes the primary computational effort. If $a$ doesn't update at iteration $i$, i.e., no candidate in $D^{(j)}$ achieves a lower loss, then in the next iteration, instead of searching over the top $k=\frac{n}{m}$, we search over the second tier of candidates, i.e., tokens from $\mathcal{V}$ with scores between top $\frac{n}{m}$ and the  top $\frac{2n}{m}$, since the top $\frac{n}{m}$ candidates were already evaluated in the previous iterations. The search proceeds methodically as described above until a better candidate is found and $a$ is updated. The whole process is described in Algorithm \ref{alg: AGGD}. 


\begin{algorithm}[h]
	\caption{Approximate Greedy Gradient Descent\ (AGGD) \label{alg: AGGD}}
	\begin{algorithmic}
		\STATE {\bfseries Input:} The token length $m$ of the adversarial passage; Initialized adversarial passage $a = [t_{1}, \cdots, t_{m}]$; Total number of iterations $N$; Size of the candidate set $n$; Search depth $d=0$.\\
  \STATE Let $k=\frac{n}{m}$ to be the per token candidate set size.
       \FOR{$j=0, 1,2, \cdots, N$ }
\FOR{$i=1, \cdots, m$}
\STATE Let $\mathcal{V}_o=\mathcal{R}(s(a))$ where $s(a)=-e_v^T\nabla \ell(a)$, $\ell(a) = -\frac{1}{|\mathcal{Q}|} \sum_{q }\text{Sim}(q, a)$ and $\mathcal{R}(\cdot)$ is a rank function. 
\STATE Let $\mathcal{X}_i = \mathcal{V}_o[(d-1)k, dk]$ be the truncated top-$(d-1)k$ to top-$dk$ candidates.
\ENDFOR
\STATE Construct candidate set at iteration $j$ as $D^{(j)}=\cup_{i\in [m]} \mathcal{X}_i$
\STATE Let $a^{'}=\arg\min_{a^{'}\in D^{(j)}}\ell(a^{'})$
\textcolor{lightgray}{// Check if a better adversarial passage exists in candidate set}
  \IF {$\ell(a^{'})<\ell(a)$}
 
    \STATE Update $a=a^{'}$, reset depth $d=0$. \textcolor{lightgray}{// Update the adversarial passage if find a better one}
    \ELSE 
   
    \STATE Update $d= d+1$  \textcolor{lightgray}{// If no update, search from top-$(d-1)k$ to top-$dk$ in next iteration}
  \ENDIF
        \ENDFOR\\
        \STATE \textbf{Return} Final adversarial passage 
        $a$
	\end{algorithmic}
\end{algorithm}

\section{Experiments}
\subsection{Experimental Details}

\textbf{Datasets} We primarily use two popular question-answering datasets: Natural Questions (NQ) \citep{kwiatkowski2019natural} and MS MARCO \citep{nguyen2016ms} for our attack. NQ, containing 132,803 question-answer pairs, is collected from Wikipedia, while MS MARCO, containing  532,761 question answer pairs, is collected from web documents. 
For in-domain attacks, we evaluate the attack on the  unseen  queries on held-out test queries of these two datasets. To test the transferability of our attack, we also evaluate it on 5 out-of-domain datasets: NFCorpus \cite{boteva2016full}, Quora, SCIDOCS \cite{cohan2020specter}, SciFact \cite{wadden2020fact}, FiQA-2018 \cite{maia201818}. These datasets contain unseen queries and corpora that are out of distribution or from entirely different domains such as biomedicine, scientific articles, and finance. Statistics of these datasets can be found in Appendix \ref{app: dataset detail}. 

\textbf{Retrievers}\quad
In our main experiments, we conduct attacks on 5 state-of-the-art retrieval models: Contriever, Contriever-MS (Contriever fine-tuned on MS MARCO) \cite{gautier2022unsupervised}, 
DPR-nq (trained on NQ), DPR-mul (trained on multiple datasets) \cite{karpukhin2020dense} and ANCE \cite{xiong2020approximate}. 

\textbf{Evaluation Metrics}\quad
After generating the adversarial passages on the training set and injecting them into the corpus, we evaluate the effectiveness of our attack using the top-$k$ \textit{attack success rate (ASR)} on test queries. Top-$k$ ASR is defined as the percentage of queries for which at least one adversarial passage is retrieved in the top-$k_r$ results, i.e.,
$\text{ASR} = \frac{1}{n_q}\sum_{i=1}^{n_q} \mathbbm{1}\{a\in \mathcal{R}_r(q^{\text{test}}_i, k_r, \mathcal{C}_{\text{test}})\}$, 
where $n_q$ is the total number of test queries and $\mathcal{R}_r(q_i^{\text{test}
}, k_r, \mathcal{C}_{\text{test}})$ is the retriever that returns the top-$k_r$ most relevant passages for the test query $q_i^{\text{test}}$. $\mathbbm{1}\{a\in \mathcal{R}_r(q_i^{\text{test}}, k_r, \mathcal{C}_{\text{test}})\}$ is the indicator function, which equals to 1 if $a\in \mathcal{R}_r(q_i^{\text{test}},  k_r, \mathcal{C}_{\text{test}})$ and 0 otherwise. A higher ASR indicates that the model is more vulnerable to attacks, and thus, the attack is more effective. We use $k_r=20$ to present our result. Since ASR depends on the size the the test corpus, to make fair comparisons across different dataset, we randomly sample $|\mathcal{C}_{\text{test}}| =10,000$ passages from the overall corpus pool. 

Additionally, we use \textit{retrieval accuracy} on the validation data, which is defined as 
\begin{equation*}
\text{RetAcc} =\frac{1}{n_{\text{val}} }\sum_{i=1}^{n_{\text{val}}} \mathbbm{1}\{\text{Sim}(q_i^{\text{val}}, p_i^{\text{val}})> \text{Sim}(q_i^{\text{val}}, a)\}
\end{equation*}
where $a$ is the adversarial passage,  $n_{\text{val}}$ is the total number of query-passage pairs in the validation set, and $p_i^{\text{val}}$ is most semantically relevant passage that should have been retrieved by query $p_i^{\text{val}}$. Lower retrieval accuracy indicates a higher chance of the model being fooled into choosing the adversarial texts $a$ over the most semantically relevant passage in the corpus. Note that the Success Rate \textbf{during training} (on validation data) can be defined through retrieval accuracy, which is simply $1-\text{RetAcc}$.

\textbf{Hyperparameters}\quad For all baselines, we use adversarial passages of $m=30$ tokens and perform the token replacement for 2000 steps. We fix the candidate set size at $n=150$. All the experiments are conducted on NVIDIA A100 GPU (with total memory 40G).

\subsection{Main Results}
\begin{table*}[t!]
\centering
\scriptsize
\begin{tabular}{ccccacc}
\toprule

\multirow{2}{*}{
\textbf{Dataset}
}&\multirow{2}{*}{
\textbf{Method}
}&\multicolumn{5}{c}{\textbf{Retriever}}\\\cline{3-7}
&& Contriever&Contriever-MS&ANCE&DPR-mul&DPR-nq\\\midrule \multirow{3}{*}{NQ} & AGGD& \textbf{92.5(2.68)} & \textbf{63.45(8.68)} & \textbf{80.92(4.82)} & \textbf{6.88(1.72)} & \textbf{2.19(0.98)}\\  & Hot Flip& 91.08(0.9) & 58.43(4.53) & 65.68(2.5) & 5.4(0.36) & 2.03 (0.25)\\  & Random& 80.24(0.92) & 32.5(2.61) & 31.0(4.3) & 3.7(0.93) & 1.66 (0.14)\\ \cline{1-7}
 \multirow{3}{*}{MS MARCO} & AGGD& \textbf{85.47(3.81)} & \textbf{24.42(17.44)} & \textbf{93.6(1.01)} & \textbf{12.79(3.86)} & 5.23 (1.01)\\  & Hot Flip& 83.72(5.93) & 22.67(12.01) & 76.16(9.21) & 9.88(1.93) & \textbf{5.81(2.01)}\\  & Random& 66.86(7.24) & 13.95(13.46) & 49.42(11.2) & 6.4(2.53) & 2.91 (1.93)\\ 
\bottomrule
\end{tabular}
\caption{In-domain attack success rate (ASR) of AGGD on NQ and MS MARCO datasets with 5 retrievers by injecting 1 adversarial passage. 
We highlight the best performing attacking method in bold (Higher ASR indicates better attack performance). Results are from 4 random runs with standard deviation in parenthesis.}
\label{tab: ASR in domain}
\vspace{-0.5cm}
\end{table*}

\textbf{In-Domain Attack}\quad We show the in-domain attack performance of AGGD and the other two baselines (HotFlip and random perturbation) in Table \ref{tab: ASR in domain}, evaluating the injection of only one adversarial passage. Our findings are as follows: (1) The pretrained Contriever model is more vulnerable to attacks. All three attack baselines achieve the highest ASR compared to other retrieval models. Poisoning with AGGD achieves 92.5\% ASR on NQ dataset and 85.47\% on MS MARCO dataset, respectively. Even using Random perturbation can achieve a relatively high ASR of 80.24\% on NQ dataset and  66.86\% on MS MARCO dataset. (2) Besides achieving comparable results in Contriever and DPR, AGGD is extremely effective in attacking ANCE, improving over the second-best baseline by 15.24\% and 17.44\% on NQ and MS MARCO, respectively. The effectiveness of AGGD in attack ANCE can also be  clearly observed during the training. Due space limitations, we provide training accuracy in Appendix \ref{appx: retrieval accuracy during training (1 adv)} (Figure \ref{fig: retriever accuracy with 1 passage}).   (3) Supervised retriever models such as DPR are more challenging to attack. Attacking DPR-nq on NQ with only 1 adversarial passage are just slightly better than random perturbation. 

\begin{table*}[t!]
\centering
\scriptsize
\begin{tabular}{cccccccc}
\toprule
\multirow{2}{*}{
\textbf{Target Domain}
}
& \multirow{2}{*}{
\textbf{Source Domain}
}& \multirow{2}{*}{
\textbf{Methods}
}& \multicolumn{5}{c}{\textbf{Retriever}}\\\cline{4-8}
&&& Contriever&Contriever-MS&ANCE&DPR-mul&DPR-nq\\\midrule
\multirow{6}{*}{FiQA-2018} & \multirow{3}{*}{NQ} & AGGD& 64.93(15.9) & 4.94(1.68) & 6.33(1.91) & 0.15(0.11) & 0.12(0.13)\\ & & Hot Flip& 69.68(8.66) & 3.05(0.6) & 4.13(0.77) & 0.46(0.55) & 0.62(0.15)\\ & & Random& 41.05(2.73) & 0.35(0.13) & 1.66(0.49) & 0.27(0.17) & 0.46(0.19)\\ & \multirow{3}{*}{MS MARCO} & AGGD& \textbf{86.54(1.88)} & \textbf{15.66(22.54)} & \textbf{17.98(1.44)} & \textbf{3.59(1.34)} & \textbf{3.36(1.79)}\\ & & Hot Flip& 78.51(6.96) & 4.59(2.89) & 11.5(0.79) & 2.16(1.13) & 3.2(1.72)\\ & & Random& 62.81(6.07) & 0.42(0.57) & 5.48(1.36) & 0.73(0.2) & 0.93(0.24)\\ \midrule
\multirow{6}{*}{NFCorpus} & \multirow{3}{*}{NQ} & AGGD& 46.83(8.1) & 17.1(4.3) & 37.69(4.2) & 8.51(1.29) & 12.23(0.46)\\ & & Hot Flip& 42.34(10.96) & 10.91(2.77) & 34.68(1.71) & 10.22(2.62) & 12.85(0.99)\\ & & Random& 24.38(5.02) & 6.58(1.43) & 18.34(4.09) & 7.43(1.3) & 11.3(0.56)\\ & \multirow{3}{*}{MS MARCO} & AGGD& \textbf{52.55(13.94)} & \textbf{18.73(16.3)} & \textbf{69.81(3.57)} & \textbf{26.16(4.51)} & \textbf{28.17(4.02)}\\ & & Hot Flip& 49.92(5.26) & 14.24(8.46) & 52.55(6.86) & 19.27(2.31) & 23.06(7.02)\\ & & Random& 30.03(6.42) & 6.42(6.45) & 42.26(6.23) & 14.47(3.59) & 14.78(1.06)\\ \midrule
\multirow{6}{*}{Quora} & \multirow{3}{*}{NQ} & AGGD& 73.92(12.16) & \textbf{25.1(7.03)} & 77.72(5.57) & 4.85(1.13) & 6.39(2.1)\\ & & Hot Flip& 78.99(3.56) & 18.92(5.84) & 72.37(2.13) & 3.36(1.02) & 7.2(1.45)\\ & & Random& 49.44(4.18) & 6.12(0.63) & 53.09(4.89) & 2.52(0.47) & 6.66(1.51)\\ & \multirow{3}{*}{MS MARCO} & AGGD& \textbf{86.92(2.49)} & 18.83(21.47) & \textbf{91.11(1.48)} & \textbf{22.29(4.24)} & \textbf{25.96(5.05)}\\ & & Hot Flip& 83.26(5.87) & 12.82(7.64) & 84.18(2.25) & 15.38(4.14) & 23.07(3.67)\\ & & Random& 62.57(6.63) & 3.81(4.12) & 74.06(0.75) & 8.82(1.14) & 12.29(3.33)\\ \midrule
\multirow{6}{*}{SCIDOCS} & \multirow{3}{*}{NQ} & AGGD& 21.05(7.78) & 11.95(7.99) & 9.78(1.97) & 0.88(0.29) & 0.38(0.08)\\ & & Hot Flip& \textbf{27.02(13.4)} & \textbf{14.32(4.48)} & 7.1(0.87) & 1.18(0.4) & 0.35(0.11)\\ & & Random& 12.88(3.76) & 0.78(0.61) & 3.05(0.93) & 0.6(0.37) & 0.3(0.07)\\ & \multirow{3}{*}{MS MARCO} & AGGD& 23.22(6.99) & 14.3(20.69) & \textbf{31.9(7.2)} & \textbf{3.88(1.62)} & \textbf{1.08(0.42)}\\ & & Hot Flip& 24.98(6.33) & 12.25(7.52) & 21.88(4.33) & 3.12(1.4) & 0.85(0.32)\\ & & Random& 15.0(1.53) & 1.27(1.83) & 10.62(5.11) & 2.48(1.38) & 0.2(0.07)\\ \midrule
\multirow{6}{*}{SciFact} & \multirow{3}{*}{NQ} & AGGD& 22.92(4.46) & 1.0(0.62) & 6.75(1.92) & - & - \\ & & Hot Flip& 24.5(12.16) & 0.92(0.72) & 3.58(0.98) & 0.17(0.29) & - \\ & & Random& 8.25(3.42) & - & 0.5(0.37) & - & - \\ & \multirow{3}{*}{MS MARCO} & AGGD& \textbf{29.08(10.43)} & \textbf{2.67(4.43)} & \textbf{30.33(6.47)} & \textbf{2.42(0.6)} & 0.5(0.29)\\ & & Hot Flip& 21.58(6.52) & 1.25(1.11) & 16.5(4.78) & 2.0(0.24) & \textbf{0.67(0.71)}\\ & & Random& 9.83(5.46) & - & 7.33(4.96) & 1.17(0.96) & 0.17(0.17)\\ 
\bottomrule
\end{tabular}
\caption{Out-of-domain top-20 attack success rate with only 1 adversarial passage. Results are averaged over 4 random runs with standard deviation shown in parenthesis. The combinations of attack and source dataset that achieve the highest ASR on each target domain and retriever are highlighted in bold.}
\label{tab: ASR out of domain}
\end{table*}

\textbf{Out-of-Domain Attack}\quad
We found that the generated adversarial passages can transfer across different domains. In Table \ref{tab: ASR out of domain}, we use adversarial passages generated from training set of NQ and MS MARCO and insert them into the corpora of retrieval tasks in other domains. We found that (1) Compared to adversarial passages generated from NQ dataset, those trained from MS MARCO generally perform better in out-of-domain attacks, possibly because MS MARCO contains more training data.  (2) Contriever models are still the most vulnerable to corpus poisoning attacks. For example, inserting a single adversarial passage generated by  AGGD into FiQA-2018 achieves a top-$k_r=20$ ASR of 86.54\%, and inserting it into Quora can trick the model into returning the adversarial passage in the top-20 retrieved passage for 86.92\% of the queries. (3) Quora are easier domains to attack. AGGD achieves over 20\% ASR even in DPR-mul and DPR-nq, which is surprising since, as shown in the in-domain attack results in Table \ref{tab: ASR in domain}, attacking DPR-mul and DPR-nq is extremely hard , even with adversarial passage trained from in-domain data. (Though NFCorpus has high attack success rate, NFCorpus also has a smaller testing corpus).

\subsection{Analysis and Ablation Study} \label{sec: ablation}

\begin{figure*}[t!]
    \centering
    \begin{subfigure}[t]{0.45\textwidth}
        \centering
        \includegraphics[height=1.2in]{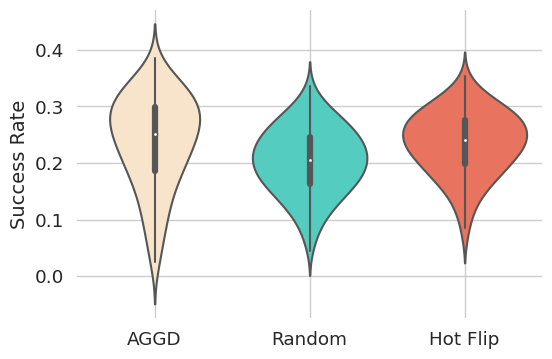}
        \caption{Attack Success rate of candidates sets collected by different methods. (Averaged over 400 candidate sets sampled.)}
         \label{fig: avg candidate succ rate}
    \end{subfigure}
    \hfill
    \begin{subfigure}[t]{0.45\textwidth}
        \centering
        \includegraphics[height=1.2in]{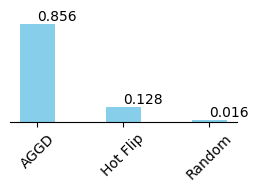}
        \caption{The proportion of times the best candidate occurs in the candidate sets collected by AGGD, HotFlip, and Random. }
        \label{fig: count}
    \end{subfigure}
    \caption{Experiments on Contriever with NQ dataset illustrate that the candidate set collected by AGGD has higher overall quality (left) and is more likely to contain the best candidate (right).}
    \label{fig: candidate set quality}
    \vspace{-0.5cm}
\end{figure*}

\textbf{Candidate Set Quality}\quad Our results indicate that candidate sets selected using AGGD have better quality than those selected using HotFlip and Random perturbation. To demonstrate this, we randomly sampled a passage and used AGGD, HotFlip and random perturbation to select their respective candidate sets with $m=150$ candidates. We then evaluated the success rate (i.e., $1-\text{RetAcc}$) on the validation set for all the candidates. The results, averaged from 400 random samples, are shown in Figure \ref{fig: avg candidate succ rate}. We can observe that the candidate set from AGGD has higher mean success rate on validation data. Additionally, the higher confidence bound in the AGGD candidate set signifies that AGGD's candidate set not only has higher overall quality, but is also more likely to contain the best candidate when compared horizontally across candidate sets from other methods. To verify this, we counted how often the best candidate occurs in candidate sets selected by these methods and found that more than 85\% of the time, the AGGD candidate set contains the best candidate, while less than $13\%$ show up in HotFlip candidate set, as shown in Figure \ref{fig: count}. More experiments on other retrievers models further verify that the AGGD candidate set has higher quality (Figure \ref{fig: additional experiments for candidate set quality} in Appendix \ref{apx: candidate set quality experiments on more dataset}).

\textbf{The Effects of Candidate Set Size $n$}\quad
\begin{figure}[h]
    \centering
    \includegraphics[width=0.5\textwidth]{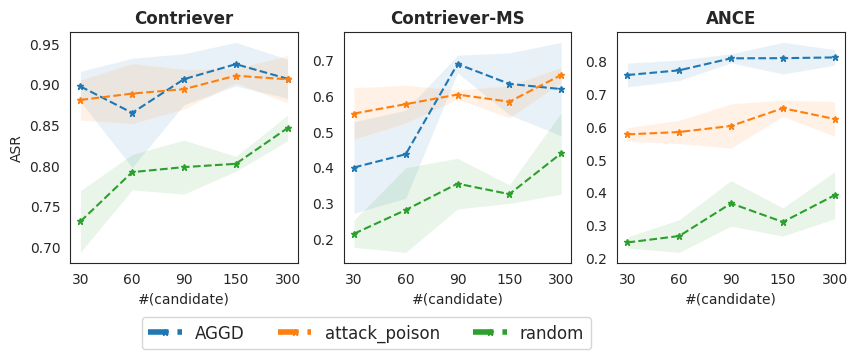}
    \caption{The effect of candidate set size $n$ on the attack success rate.}
    \label{fig: varying num candidate(main)}
\end{figure}
In Figure \ref{fig: varying num candidate(main)}, we show how the candidate set size $n$ affects the ASR evaluated on NQ dataset with three retrievers. Generally, we can observe that increasing size of the candidate set improves the ASR. The effect of the candidate set size would be more pronounced when a larger range of candidate set sizes is considered.

\begin{table*}[h]
\centering
\scriptsize
\begin{tabular}{|c|c|c|c|c|c|c|c|}
\hline
 & LLaMa-2-7B & LLaMa-2-13B & Vicuna-7B &Vicuna-13B& Vicuna-33B & GPT-3.5 & GPT-4 \\ \hline
\multicolumn{8}{|c|}{\textbf{MS MARCO}} \\ \hline
AGGD & \textbf{0.81(0.00)} & \textbf{0.79(0.01)} & \textbf{0.73(0.01)} & 0.72(0.01) & 0.67(0.01) & \textbf{0.80(0.01)} & \textbf{0.88(0.01)} \\ \hline
HotFlip & 0.80(0.02) & 0.77(0.02) & 0.69(0.02) & \textbf{0.73(0.04)} & \textbf{0.69(0.02)} & 0.79(0.01) & 0.86(0.01) \\ \hline
\multicolumn{8}{|c|}{\textbf{NQ}} \\ \hline
AGGD & \textbf{0.85(0.00)} & 0.94(0.00) & \textbf{0.81(0.00)} & \textbf{0.82(0.00)} & \textbf{0.68(0.00)} & \textbf{0.92(0.00)} & \textbf{0.97(0.00)} \\ \hline
HotFlip & 0.83(0.00) & \textbf{0.95(0.00)} & 0.80(0.00) & 0.81(0.00) & 0.67(0.00) & 0.87(0.01) & 0.96(0.01) \\ \hline
\end{tabular}
\caption{Comparing Attack success rate of PoisonedRAG (white-box) using HotFlip and AGGD to find the adversarial texts. }
\label{table: extending to RAG}
\vspace{-0.5cm}
\end{table*}

\textbf{The effect of the token length $m$}\quad
\begin{figure}[h]
    \centering
    \includegraphics[width=0.5\textwidth]{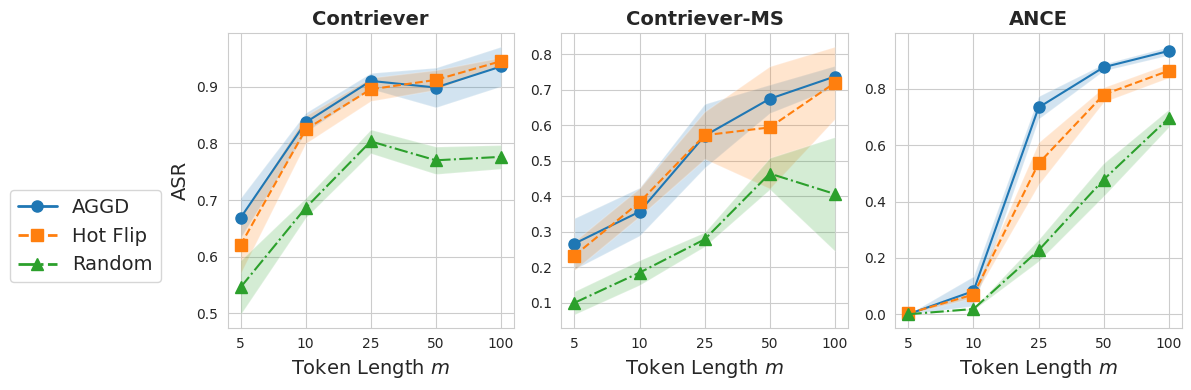}
    \caption{The effect of the number of tokens, with fixed candidate set size $n=100$ and varying adversarial passage token length $m=\{5, 10, 25, 50, 100\}$. }
    \label{fig: varying token length(main)}
    \vspace{-0.5cm}
\end{figure}
In Figure \ref{fig: varying token length(main)}, we show the attack success rate for varying token lengths  $m=\{5, 10, 25, 50, 100\}$ while fixing the candidate set size $n=100$. Generally, we find that larger token lengths result in higher attack success rate upon convergence. This is intuitive since larger $m$ indicates a larger subspace in the continuous dense embedding space, thus providing a higher \textit{lower bound} on the effectiveness of the adversarial passage we can find.

\subsection{Extending  AGGD to Knowledge Poisoning Attacks}
Similar to Hot Flip, which can be used for many other adversarial attacks such as adversarial prompt generation \cite{zou2023universal} and knowledge poisoning in RAG \cite{zou2024poisonedrag}. AGGD can also be conveniently used as a plug-in replacement for HotFlip in these attacks. In Table \ref{table: extending to RAG}, we experiment with PoisonedRAG \cite{zou2024poisonedrag} in their white box setting, where Hot Flip was originally used to craft adversarial texts. The details in reproducing Table \ref{table: extending to RAG} are given in Appendix \ref{apdx: attacking RAG}.  
We find that AGGD achieves a comparable or better attack success rate than Hot Flip in both MS MARCO and NQ dataset on multiple LLMs.

\section{Conclusion}
In this paper, we propose a new adversarial attack method called AGGD, a gradient-based search algorithm that systematically and structurally finds potential perturbations to optimize the objective function. We use the corpus poisoning attacks as the main example to demonstrate the effectiveness of our algorithm. Experiments on multiple datasets and retrievers show that the proposed approach is effective in corpus poisoning attacks, achieving high attack success rate in both in-domain and out-of-domain scenarios, even with an extremely low poison rate. Additional experiments on other adversarial attacks indicate the potential of AGGD as a competitive alternative to the widely used HotFlip.

\section{Limitations and Future Work}
While we use corpus poisoning attacks to showcase our AGGD algorithm, the proposed attack framework is versatile and applicable to a wide range of adversarial attack scenarios, as many of these can be formulated as discrete optimization problems. However, the generated output sequences often lack semantic coherence, making the adversarial corpus easily detectable and filterable. A promising direction for future work is to reformulate the problem as a constrained optimization problem, focusing on producing semantically meaningful adversarial passages that are more difficult to defend against, even if this may compromise the overall attack success rate. 
\section{Ethics Statement}
Our work studies the vulnerability of dense retriever and corpus poison attack. The propose attack AGGD shows higher attach success rate especially for ANCE model, compared to previous HotFlip attack, which could be used for spread garbage information. Future research based on this paper should be exercised with caution and consider the potential consequences. 

\bibliography{custom}

\appendix
\onecolumn
\section{Additional Experiments}

\subsection{Comparing HotFlip with Random Perturbation} \label{apx: motivational experiments}

\begin{figure*}[h]
    \centering
    \includegraphics[width=1\textwidth]{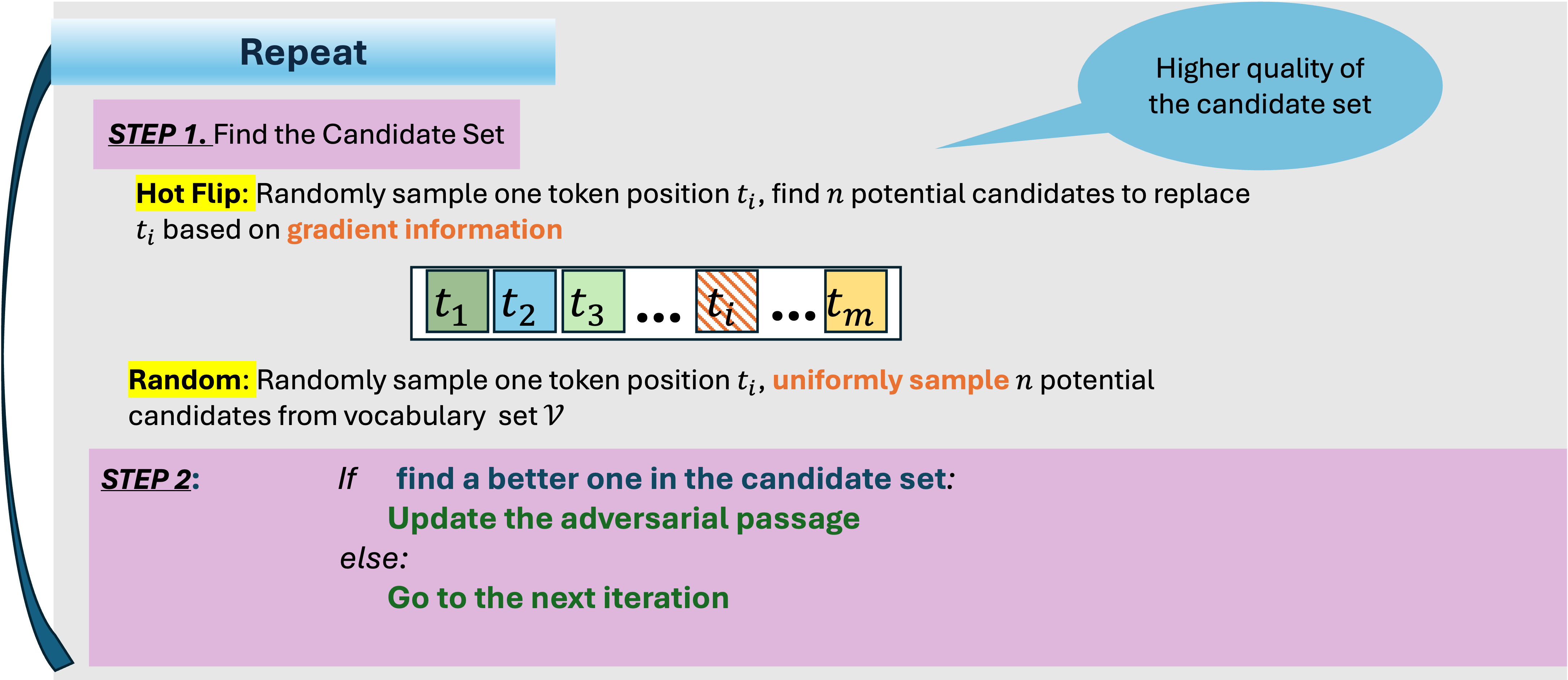}
    \caption{Comparing HotFlip attack and the random perturbation attack. The main difference is how they select the candidate set. HotFlip uses the gradient information to enhance the quality of the candidate set, whereas random perturbation selects candidates uniformly from $\mathcal{V}$ without leveraging gradient information.}
    \label{fig: intro}
\end{figure*}

\begin{figure*}[tbh]
    \centering
    \includegraphics[width=0.7\textwidth]{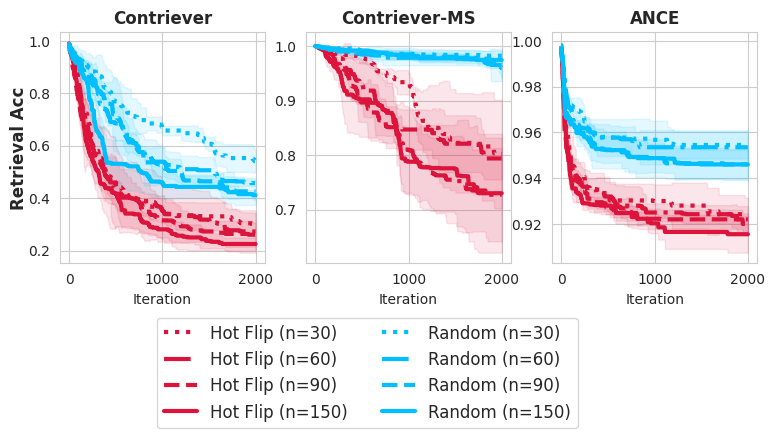}
    \caption{Retrieval accuracy of HotFlip and random perturbation on the NQ dataset with varying candidate set size $n=\{30, 60, 90, 150\}$. \textbf{Lower} retrieval accuracy suggests a more successful adversarial passage attack, as evaluated on the validation set during greedy search iterations.}
    \label{fig: compare hotflip with random}
\end{figure*}

\begin{figure*}[h]
    \centering
    \includegraphics[width=0.7\textwidth]{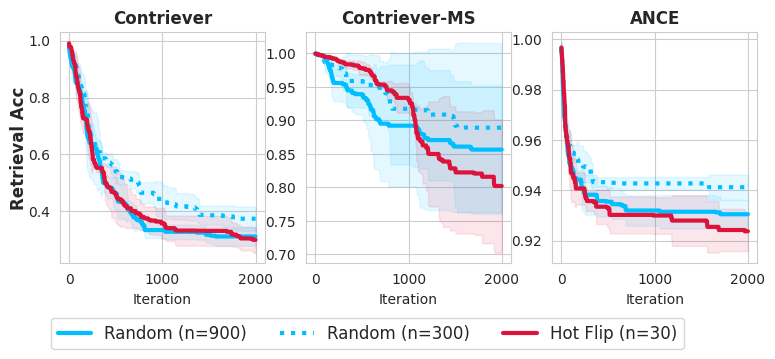}
    \caption{Retrieval accuracy of HotFlip and random perturbation. Hot Flip only needs to maintain a candidate set of size $n=30$ to achieve the same performance of random perturbation with a candidate set size $n=900$. }
    \label{fig: one candidate}
\end{figure*}

As illustrated in Figure \ref{fig: intro}, HotFlip first randomly selects a token position for replacement and identifies $n$ potential alternatives for that position. It updates the adversarial passage with the most effective perturbation from this candidate set if it improves attack performance. If no perturbation enhances performance, HotFlip progresses to the next iteration by randomly selecting another token position and repeating the process.
It is useful then to compare Hotflip vs.\ a natural black box \texttt{Random} approach to candidate generation. As  depicted in Figure \ref{fig: intro}), the Random approach creates a candidate  set by randomly sampling from the vocabulary. 
\paragraph{HotFlip is more Efficient Than Random Perturbation}
In experiments using three dense retrievers on the widely used Natural Questions (NQ) dataset, we find that HotFlip outperforms the Random perturbation baseline, as shown in Figure~\ref{fig: compare hotflip with random}. Specifically,  HotFlip consistently achieves low retrieval accuracy on the validation set while maintaining candidate sets of the same size \textit{or smaller}. (Red lines are mostly lower than the blue lines.)
  Equivalently, to achieve the same degree of attack effect, HotFlip is more efficient because it requires a smaller  candidate set. For example, as depicted in Figure \ref{fig: one candidate}, Random perturbation needs to maintain a candidate set of size 
$n=900$ to match the performance of HotFlip with only 
$n=30$ candidates. This means that the size of Random perturbation candidate set is 30 times larger and consequently,  Random perturbation is 30 times slower.

 HotFlip and Random perturbation share the same intuition of maintaining a candidate set and then take a greedy token swap based on this set. Though many factors can influence the  efficiency and the effectiveness the greedy search, one of the major factors is the quality of the candidate set. If the candidate set is of high quality, we can maintain a smaller size candidate set, reducing the number of searches at each greedy step. For example, HotFlip filters down the most likely potential tokens from $|\mathcal{V}|$ to $n$ based on the gradient information, while random perturbation samples $n$ tokens from $\mathcal{V}$. As a result, random perturbation has a lower quality candidate set. 
Therefore, it is crucial to improve the quality of the candidate set.

\subsection{Retrieval Accuracy during Training} \label{appx: retrieval accuracy during training (1 adv)}
\begin{figure*}[h]
    \centering
    \includegraphics[width=1\textwidth]{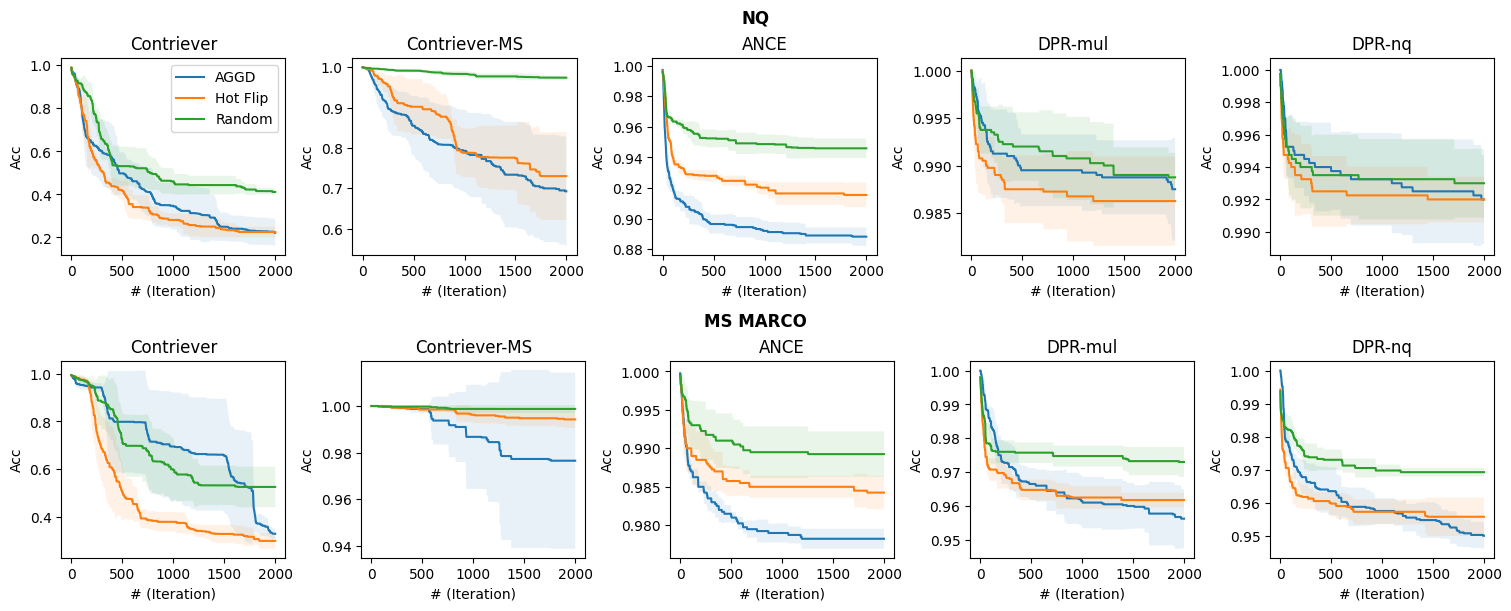}
    \caption{Retrieval accuracy (the portion of the queries that have higher similarity to the adversarial passage than the gold passage) on validation data during training. Lower retrieval accuracy indicates higher chance of adversarial passage being retrieved.}
    \label{fig: retriever accuracy with 1 passage}
\end{figure*}

\paragraph{Retrieval Accuracy for 1 Adversarial Passage}
In Figure \ref{fig: retriever accuracy with 1 passage}, we plot the Retrieval accuracy during training on both NQ and MS MARCO datasets for all 5 retrievers as a complement to the results in Table \ref{tab: ASR in domain}. During training, at each iteration, we evaluate the retrieval accuracy of the best adversarial passage $a_i$ from the candidate set, and compare it with the current adversarial passage to decide whether to update it. Due to the greedy nature of these attacks, Retrieval accuracy is guaranteed to descend. Figure \ref{fig: retriever accuracy with 1 passage} shows that, for both NQ and MS MARCO dataset, the harder-to-attack models such as ANCE, DPR-mul and DPR-nq converge with high retrieval accuracy. In contrast, for the easier-to-attack models such as Contriever, the attack methods have not yet converged, even though the retrieval accuracy is as low as around 0.2.

\paragraph{Effect of the Candidate Set Size to Retrieval Accuracy}
\begin{figure*}[h]
    \centering
    \includegraphics[width=1\textwidth]{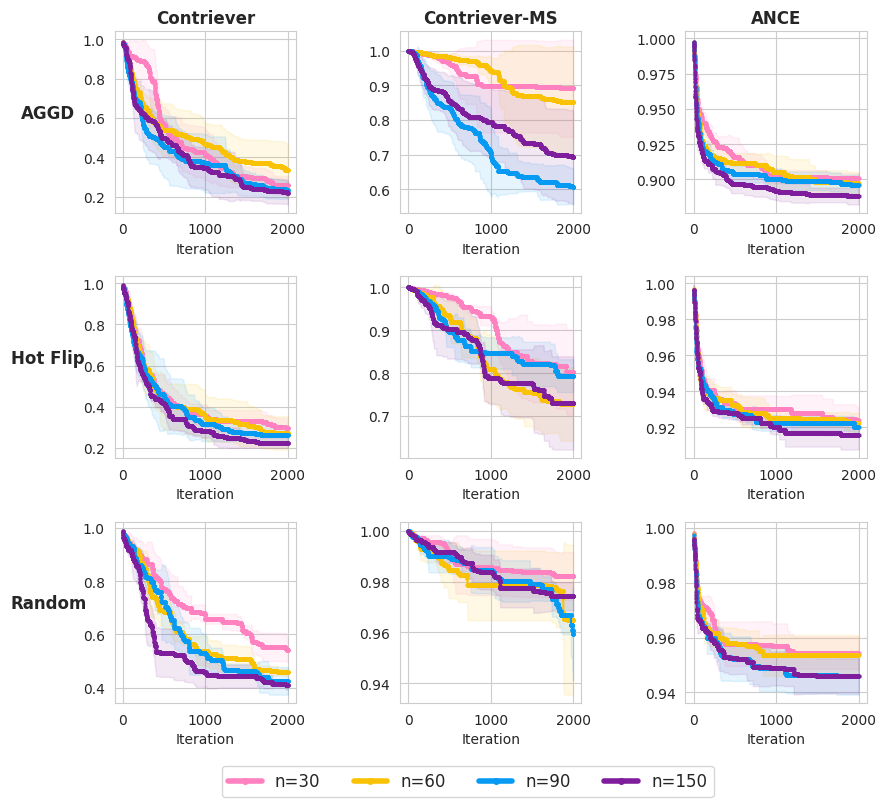}
    \caption{The retrieval accuracy during training with various candidate set sizes. ($n=30, 60, 90, 150)$. }
    \label{fig: varying num candidate}
\end{figure*}

In Figure \ref{fig: varying num candidate}, we illustrate the retrieval accuracy during training with varying candidate set sizes on NQ dataset and 3 retrievers. Larger candidate set sizes generally lead to lower retrieval accuracy. 

\begin{figure*}[h]
    \centering
    \includegraphics[width=1\textwidth]{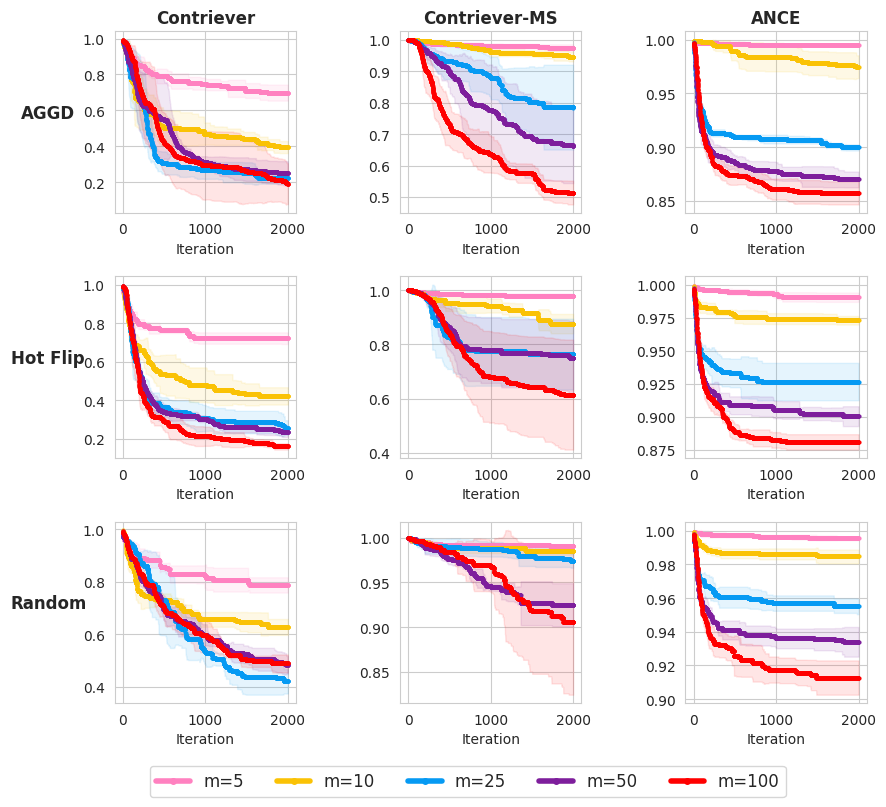}
    \caption{The retrieval accuracy during training with various token lengths ($m\in\{5, 10, 25, 50, 100\}$) and a fixed candidate set size $n=100$. }
    \label{fig: varying token length}
\end{figure*}

\paragraph{Effect of Token Length on Retrieval Accuracy}
In Figure \ref{fig: varying token length}, we show the training retrieval accuracy with various token length settings. We find that, within a proper range, longer token lengths lead to more effective adversarial passages. 

\section{Experimental Results with 10 Adversarial Passages}
To generate multiple adversarial passages, we follow \cite{zhong2023poisoning} by clustering similar queries based on their embeddings using the $k$-means algorithm. We then generate one adversarial passage for each cluster.

\subsection{Candidate Set Quality} 
\label{apx: candidate set quality experiments on more dataset}
\begin{figure*}[t!]
    \centering
    \begin{subfigure}[t]{0.6\linewidth}
        \centering
        \includegraphics[width=\linewidth]{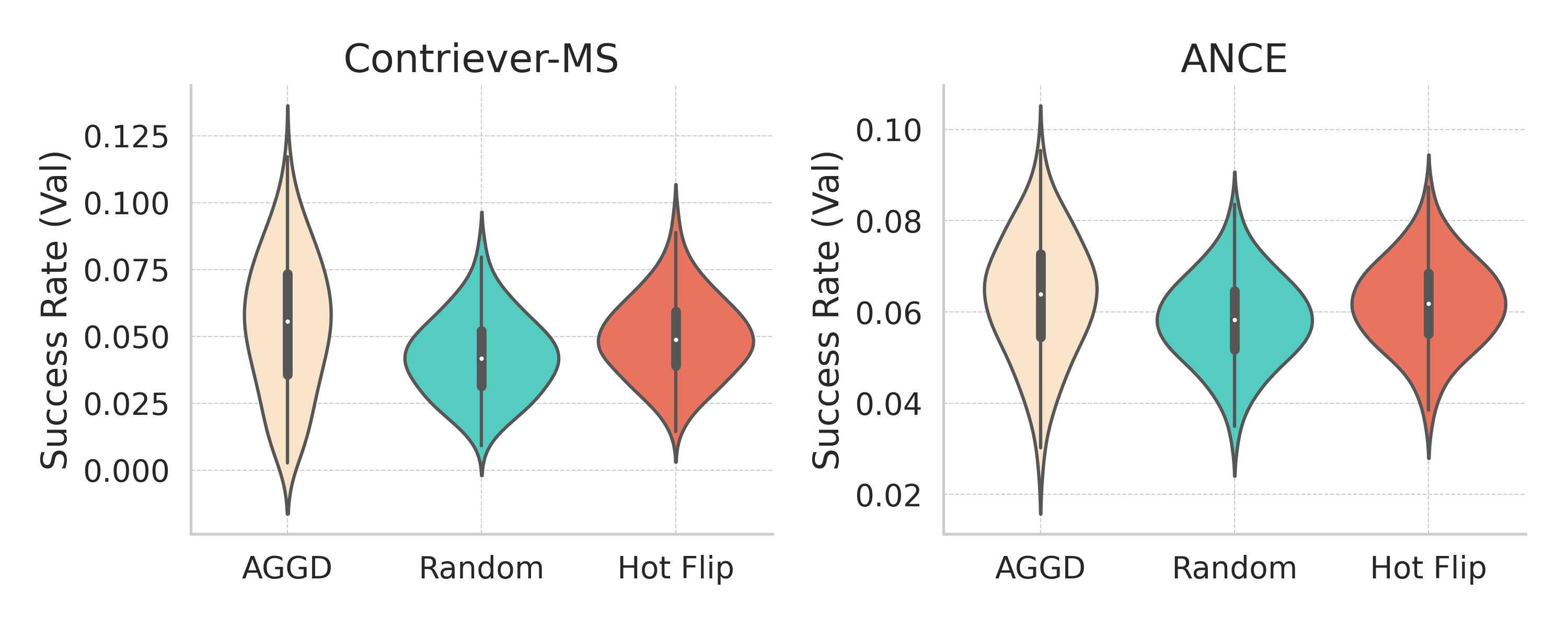}
        \caption{Attack Success rate of candidate sets collected by different methods. (Results are averaged over 400 candidate sets sampled when training with Contriever-MS (left) and ANCE (right) on NQ dataset.)}
         \label{fig: candidate quality appendix violinplot}
    \end{subfigure}%
    \hfill
    \begin{subfigure}[t]{0.6\linewidth}
        \centering
        \includegraphics[width=\linewidth]{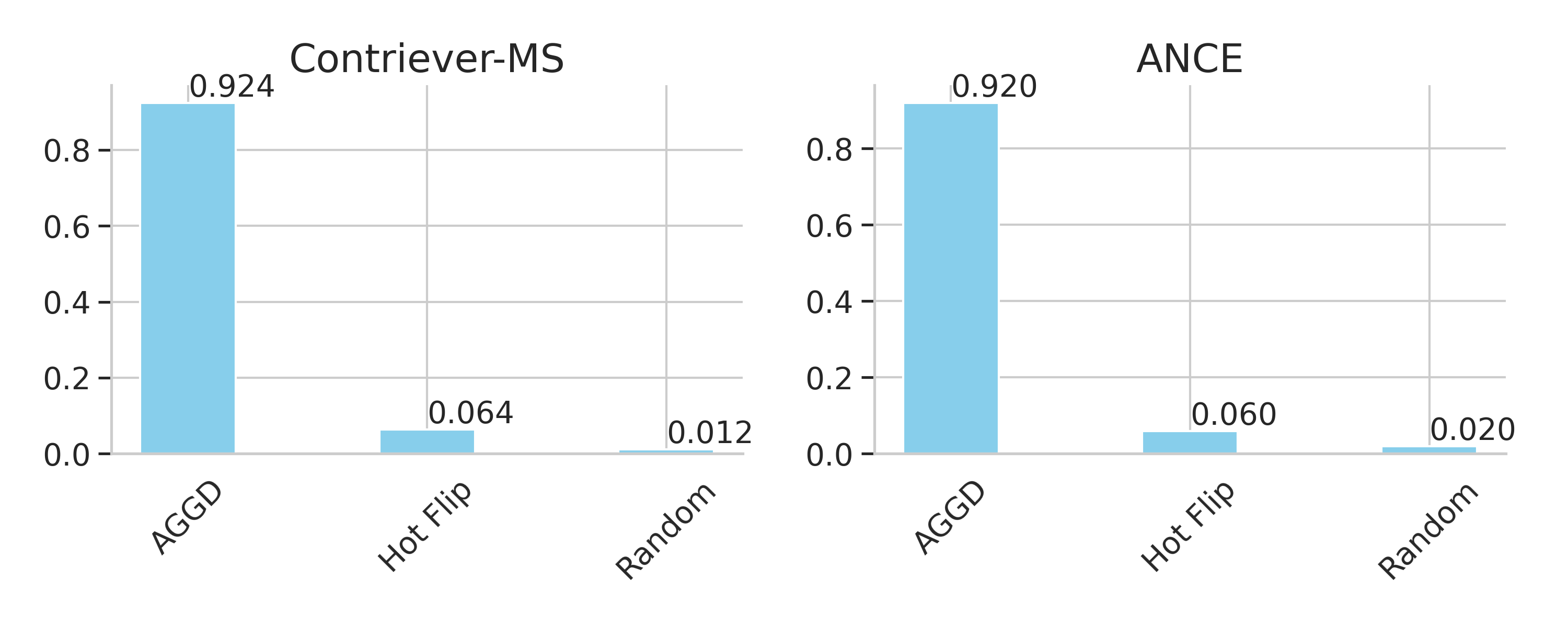}
        \caption{The portion of the times the best candidate occurs in the candidate sets collected by AGGD, HotFlip and Random, respectively, when training on NQ dataset with Contriever-MS (left) and ANCE (right).}
        \label{fig: candidate quality appendix barplot}
    \end{subfigure}
    \caption{Additional experiments on Contriever-MS and ANCE with NQ dataset illustrate that candidate sets collected by AGGD (a) has higher overall quality and (b) are more likely to contain the best candidate. }
    \label{fig: additional experiments for candidate set quality}
\end{figure*}

In Figure \ref{fig: additional experiments for candidate set quality}, we present additional experiments with two more retriever models: Contriever-MS and ANCE. We observed a similar trend, over the 400 random samples, with more than 92\% of the best candidate belonging to AGGD candidate set. This further supports the conclusions in Section \ref{sec: ablation}.

\paragraph{In-Domain Attack with 10 Adversarial Passages}
Table \ref{tab: ASR in domain (10 adv)} shows the in-domain attack results of inserting 10 adversarial passages into the NQ and MS MARCO datasets.  Similar to injecting only 1 adversarial passage, we found that (1)
The pretrained Contriever model is easy to attack: even with just 10 adversarial passages, all three baselines (AGGD, HotFlip and Random perturbation)  successfully attack more than 80\% of the queries in both NQ and MS MARCO datasets, tricking the Contriever into returning the adversarial passage among the top-20 retrieved results. For NQ dataset, even random perturbation achieves a high ASR of 80\%, while the other two methods using gradient information achieves higher ASR of $> 91\%$. (2) AGGD still outperforms HotFlip on ANCE, with improvements of 18.62\% and 25.58\% on NQ and MS MARCO datasets, respectively. 
\begin{table*}[t!]
\centering
\scriptsize
\begin{tabular}{ccccacc}
\toprule
\multirow{2}{*}{
\textbf{Dataset}
}&\multirow{2}{*}{
\textbf{Methods}
}&\multicolumn{5}{c}{\textbf{Retriever}}\\\cline{3-7}
& &Contriever&Contriever-MS&ANCE&DPR-mul&DPR-nq\\\midrule
\multirow{3}{*}{NQ} & AGGD& 91.57(3.5) & \textbf{69.42(1.38)} & \textbf{82.08(2.01)} & \textbf{8.23(1.48)} & 1.9 (0.07)\\  & Hot Flip& \textbf{91.6(0.87)} & 57.37(5.95) & 63.46(1.55) & 5.08(0.1) & \textbf{2.19(0.25)}\\  & Random& 80.49(0.83) & 33.73(2.27) & 27.3(2.74) & 3.94(1.22) & 1.65 (0.06)\\ \hline
 \multirow{3}{*}{MS MARCO} & AGGD& 83.72(4.65) & \textbf{13.95(13.95)} & \textbf{93.02(0.0)} & \textbf{11.63(2.33)} & 5.81 (1.16)\\  & Hot Flip& \textbf{88.37(2.33)} & 13.95(11.63) & 67.44(2.33) & 10.46(1.16) & \textbf{6.98(2.33)}\\  & Random& 65.12(9.3) & 1.16(1.16) & 52.33(15.12) & 8.14(1.16) & 3.49 (1.16)\\ \hline
\bottomrule
\end{tabular}
\caption{In-domain attack success rate (ASR) of AGGD on NQ and MS MARCO dataset with 5 retrievers by injecting 10 adversarial passages).
The best-performing attacking method is highlighted with in bold (higher ASR indicates better attack performance). }
\label{tab: ASR in domain (10 adv)}
\end{table*}
\paragraph{Out-of-Domain Attack with 10 Adversarial Passages}
In Table \ref{tab: ASR out of domain(10 adversarial passage)}, we perform the out-of-domain attack by inserting more adversarial passages. 
We find that, inserting more adversarial passages significantly improves the attack transferability of HotFlip, enabling it to outperform AGGD in models such as Contriever-MS, DPR-mul and DPR-nq. However, AGGD still performs much better than HotFlip when using ANCE as the retriever.

\begin{table*}[t!]
\centering
\scriptsize
\begin{tabular}{cccccccc}
\toprule
\multirow{2}{*}{
\textbf{Target Domain}
}
& \multirow{2}{*}{
\textbf{Source Domain}
}& \multirow{2}{*}{
\textbf{Methods}
}& \multicolumn{5}{c}{\textbf{Retriever}}\\\cline{4-8}
&&& Contriever&Contriever-MS&ANCE&DPR-mul&DPR-nq\\\midrule
\multirow{6}{*}{FiQA-2018} & \multirow{3}{*}{NQ} & AGGD& 92.67(0.23) & 30.86(7.87) & 19.52(1.47) & 6.02(0.15) & 4.71(1.47)\\ & & Hot Flip& 84.1(0.77) & 24.23(2.31) & 13.35(2.7) & 3.94(2.55) & 3.24(1.23)\\ & & Random& 83.1(0.39) & 17.82(5.32) & 8.26(0.54) & 2.7(0.23) & 2.31(0.46)\\ & \multirow{3}{*}{MS MARCO} & AGGD& \textbf{97.69(0.31)} & 50.54(25.23) & \textbf{37.19(0.31)} & \textbf{27.31(7.87)} & 20.6(7.64)\\ & & Hot Flip& 95.22(1.08) & \textbf{56.48(3.4)} & 28.16(2.08) & 19.37(6.4) & \textbf{28.47(1.31)}\\ & & Random& 86.73(0.15) & 6.94(0.15) & 13.27(1.39) & 15.66(7.48) & 7.25(4.63)\\ \midrule
\multirow{6}{*}{NFCorpus} & \multirow{3}{*}{NQ} & AGGD& 78.02(1.24) & 35.76(3.25) & 64.55(3.87) & 18.11(4.8) & 18.89(3.1)\\ & & Hot Flip& 85.45(0.31) & 37.0(3.87) & 48.76(1.08) & 20.43(0.31) & 20.59(1.39)\\ & & Random& 63.78(1.86) & 24.46(3.41) & 35.6(0.62) & 16.72(1.55) & 15.79(1.55)\\ & \multirow{3}{*}{MS MARCO} & AGGD& \textbf{95.36(0.62)} & 57.9(26.01) & \textbf{88.39(0.15)} & 57.89(8.05) & 52.01(3.72)\\ & & Hot Flip& 87.62(5.88) & \textbf{70.28(12.69)} & 80.34(1.39) & \textbf{59.75(10.84)} & \textbf{66.72(4.8)}\\ & & Random& 85.45(3.1) & 51.39(5.57) & 65.17(2.63) & 34.98(1.24) & 52.48(8.2)\\ \midrule
\multirow{6}{*}{Quora} & \multirow{3}{*}{NQ} & AGGD& 96.18(0.03) & 63.14(3.93) & 91.52(0.89) & 20.96(3.63) & 25.8(5.81)\\ & & Hot Flip& 92.22(1.08) & 57.88(10.45) & 84.9(2.27) & 24.82(0.15) & 32.19(1.7)\\ & & Random& 87.85(2.18) & 46.28(7.64) & 81.72(0.25) & 15.84(0.97) & 22.28(3.28)\\ & \multirow{3}{*}{MS MARCO} & AGGD& 97.87(0.93) & 61.72(7.29) & \textbf{94.52(0.17)} & \textbf{45.92(1.0)} & 51.5(3.35)\\ & & Hot Flip& \textbf{98.13(0.79)} & \textbf{69.88(3.63)} & 91.5(0.16) & 43.86(2.64) & \textbf{51.6(3.73)}\\ & & Random& 83.97(3.6) & 35.24(4.43) & 85.48(2.07) & 32.88(0.31) & 38.36(6.64)\\ \midrule
\multirow{6}{*}{SCIDOCS} & \multirow{3}{*}{NQ} & AGGD& 53.75(7.05) & 46.4(17.7) & 33.75(1.25) & 1.7(0.3) & 0.45(0.05)\\ & & Hot Flip& 49.75(2.45) & 40.7(5.4) & 26.45(6.05) & 3.35(0.15) & 0.95(0.15)\\ & & Random& 37.8(4.3) & 18.25(1.75) & 13.3(0.9) & 2.3(0.4) & 0.6(0.2)\\ & \multirow{3}{*}{MS MARCO} & AGGD& \textbf{65.5(2.6)} & 41.65(9.85) & \textbf{46.2(1.9)} & \textbf{24.9(2.2)} & 6.7(0.2)\\ & & Hot Flip& 51.45(4.75) & \textbf{68.1(8.3)} & 36.95(0.35) & 22.4(2.9) & \textbf{8.3(1.1)}\\ & & Random& 39.2(4.5) & 13.9(4.3) & 20.9(0.6) & 15.55(3.35) & 2.45(1.05)\\ \midrule
\multirow{6}{*}{SciFact} & \multirow{3}{*}{NQ} & AGGD& 60.33(6.0) & 12.5(7.17) & 39.33(9.33) & 0.5(0.5) & 0.17(0.17)\\ & & Hot Flip& 58.83(12.83) & 11.83(2.17) & 16.5(3.5) & 1.67(0.67) & 0.5(0.17)\\ & & Random& 34.33(4.33) & 3.17(2.17) & 8.17(1.5) & 0.33(0.0) & 0.5(0.17)\\ & \multirow{3}{*}{MS MARCO} & AGGD& \textbf{91.5(4.5)} & \textbf{41.5(8.5)} & \textbf{58.17(0.83)} & 12.0(5.0) & 12.0(7.67)\\ & & Hot Flip& 85.83(5.17) & 29.17(9.5) & 42.33(4.0) & \textbf{15.17(0.5)} & \textbf{14.67(7.33)}\\ & & Random& 75.5(9.5) & 8.0(1.33) & 21.5(6.5) & 5.67(1.67) & 10.33(1.0)\\
\bottomrule
\end{tabular}
\caption{Out-of-domain top-20 attack success rate with 10 adversarial passages. Due to the computational constraints, results are averaged over 2 random runs, with standard deviations shown in parentheses. The combinations of attack and source dataset that achieve the highest ASR for each target domain and retriever are highlighted in bold. }
\label{tab: ASR out of domain(10 adversarial passage)}
\end{table*}

\subsection{In-domain attack success rate with different $k$. }
\begin{table*}[t!]
\centering
\scriptsize
\begin{tabular}{cccccccc}
\toprule
&\multirow{2}{*}{
\textbf{Dataset}
}&\multirow{2}{*}{
\textbf{Method}
}&\multicolumn{5}{c}{\textbf{Retriever}}\\\cline{4-8}
&&& Contriever&Contriever-MS&ANCE&DPR-mul&DPR-nq\\\midrule 
\multirow{6}{*}{Top-1} & \multirow{3}{*}{NQ} & AGGD& \textbf{85.06(5.06)} & \textbf{44.84(12.65)} & \textbf{16.36(1.52)} & \textbf{0.39(0.15)} & - \\ & & Hot Flip& 83.77(1.43) & 39.74(7.81) & 8.66(0.62) & 0.21(0.08) & 0.03 (0.02)\\ & & Random& 66.64(2.61) & 8.82(0.87) & 1.89(0.5) & 0.07(0.04) & \textbf{0.04(0.03)}\\ \cline{2-8}
& \multirow{3}{*}{MS MARCO} & AGGD& 69.19(6.65) & 11.05(12.67) & \textbf{24.42(7.45)} & 1.16(1.16) & \textbf{2.91(1.93)}\\ & & Hot Flip& \textbf{71.51(4.15)} & \textbf{12.79(8.47)} & 10.47(2.01) & \textbf{2.33(1.64)} & 0.58 (1.01)\\ & & Random& 52.33(6.26) & 1.74(1.93) & 5.23(4.47) & - & 0.58 (1.01)\\ \hline
\multirow{6}{*}{Top-5} & \multirow{3}{*}{NQ} & AGGD& \textbf{89.84(3.53)} & \textbf{55.63(10.96)} & \textbf{49.56(3.78)} & \textbf{1.9(0.55)} & 0.27 (0.16)\\ & & Hot Flip& 88.69(0.72) & 50.64(5.96) & 31.36(1.16) & 1.29(0.26) & \textbf{0.28(0.1)}\\ & & Random& 75.28(1.8) & 20.07(1.44) & 10.12(1.82) & 0.71(0.28) & 0.24 (0.11)\\ \cline{2-8}
& \multirow{3}{*}{MS MARCO} & AGGD& \textbf{81.98(5.03)} & \textbf{20.35(16.72)} & \textbf{66.28(5.33)} & \textbf{8.72(3.02)} & \textbf{3.49(1.16)}\\ & & Hot Flip& 80.23(6.04) & 19.77(10.2) & 36.05(9.08) & 3.49(1.16) & 2.33 (0.0)\\ & & Random& 62.21(7.6) & 8.14(7.07) & 20.93(4.03) & 3.49(2.01) & 0.58 (1.01)\\ \hline
\multirow{6}{*}{Top-10} & \multirow{3}{*}{NQ} & AGGD& \textbf{91.28(3.03)} & \textbf{59.39(9.83)} & \textbf{66.43(4.42)} & \textbf{3.62(1.02)} & \textbf{0.82(0.38)}\\ & & Hot Flip& 89.85(0.82) & 54.52(5.41) & 47.55(1.73) & 2.69(0.24) & 0.72 (0.2)\\ & & Random& 77.8(1.45) & 25.91(2.07) & 18.8(3.26) & 1.61(0.45) & 0.66 (0.1)\\ \cline{2-8}
& \multirow{3}{*}{MS MARCO} & AGGD& \textbf{84.88(3.86)} & \textbf{22.67(17.35)} & \textbf{85.47(2.53)} & \textbf{9.3(3.68)} & \textbf{4.07(1.01)}\\ & & Hot Flip& 81.4(5.45) & 20.35(10.58) & 57.56(9.36) & 4.65(1.64) & 2.91 (1.01)\\ & & Random& 65.12(7.89) & 11.63(10.78) & 35.47(8.28) & 4.65(2.85) & 1.16 (1.16)\\ \hline
\multirow{6}{*}{Top-20} & \multirow{3}{*}{NQ} & AGGD& \textbf{92.5(2.68)} & \textbf{63.45(8.68)} & \textbf{80.92(4.82)} & \textbf{6.88(1.72)} & \textbf{2.19(0.98)}\\ & & Hot Flip& 91.08(0.9) & 58.43(4.53) & 65.68(2.5) & 5.4(0.36) & 2.03 (0.25)\\ & & Random& 80.24(0.92) & 32.5(2.61) & 31.0(4.3) & 3.7(0.93) & 1.66 (0.14)\\ \cline{2-8}
& \multirow{3}{*}{MS MARCO} & AGGD& \textbf{85.47(3.81)} & \textbf{24.42(17.44)} & \textbf{93.6(1.01)} & \textbf{12.79(3.86)} & 5.23 (1.01)\\ & & Hot Flip& 83.72(5.93) & 22.67(12.01) & 76.16(9.21) & 9.88(1.93) & \textbf{5.81(2.01)}\\ & & Random& 66.86(7.24) & 13.95(13.46) & 49.42(11.2) & 6.4(2.53) & 2.91 (1.93)\\ \hline
\multirow{6}{*}{Top-50} & \multirow{3}{*}{NQ} & AGGD& \textbf{93.87(2.26)} & \textbf{68.69(6.66)} & \textbf{93.7(3.21)} & \textbf{15.59(3.41)} & \textbf{7.31(2.02)}\\ & & Hot Flip& 92.56(0.82) & 63.28(3.64) & 85.93(2.4) & 13.37(1.1) & 6.84 (1.1)\\ & & Random& 83.01(0.46) & 42.35(3.55) & 53.01(5.31) & 9.6(1.76) & 5.46 (0.51)\\ \cline{2-8}
& \multirow{3}{*}{MS MARCO} & AGGD& \textbf{87.21(5.33)} & 31.98(20.3) & \textbf{99.42(1.01)} & \textbf{22.67(5.55)} & \textbf{13.37(3.02)}\\ & & Hot Flip& 84.88(4.19) & \textbf{32.56(18.16)} & 95.35(1.64) & 19.77(3.49) & 10.46 (5.07)\\ & & Random& 69.19(4.47) & 19.19(17.35) & 75.58(5.33) & 12.79(3.49) & 4.65 (1.64)\\ \hline
\multirow{6}{*}{Top-100} & \multirow{3}{*}{NQ} & AGGD& \textbf{94.77(2.08)} & \textbf{72.59(5.28)} & \textbf{98.2(1.31)} & \textbf{27.94(5.7)} & \textbf{18.0(3.6)}\\ & & Hot Flip& 93.58(0.66) & 67.27(2.97) & 94.94(1.63) & 24.48(2.26) & 16.8 (1.79)\\ & & Random& 85.11(0.41) & 50.86(4.34) & 71.57(5.81) & 18.6(3.15) & 12.81 (1.34)\\ \cline{2-8}
& \multirow{3}{*}{MS MARCO} & AGGD& \textbf{89.54(4.79)} & \textbf{40.12(24.36)} & \textbf{100.0(0.0)} & \textbf{41.28(5.03)} & \textbf{20.93(4.35)}\\ & & Hot Flip& 86.05(4.93) & \textbf{40.12(18.12)} & \textbf{100.0(0.0)} & 31.98(5.04) & 18.02 (5.78)\\ & & Random& 70.93(2.6) & 23.26(18.09) & 90.12(6.01) & 18.02(2.53) & 9.88 (4.47)\\ \hline
\end{tabular}
\caption{In-domain attack success rate (ASR) of AGGD on NQ and MS MARCO datasets with 5 retrievers by injecting 1 adversarial passage with varying $k =\{1, 5, 10, 20, 50, 100\}$. Results are from 4 random runs with standard deviation in parenthesis.For the ease of presentation, we omit the results through `-' if top-$k$ ASR is smaller than 0.1\%.}
\label{tab: ASR in domain; different k}
\end{table*}

\subsection{Transferability over different retrievers.}
\begin{table*}[t!]
\centering
\scriptsize
\begin{tabular}{ccccccc}
\toprule
\multirow{2}{*}{
\textbf{Source Retriever}
}&\multirow{2}{*}{
\textbf{Method}
}&\multicolumn{5}{c}{\textbf{Target Retriever}}\\\cline{3-7}
&& Contriever&Contriever-MS&ANCE&DPR-mul&DPR-nq\\\midrule 
\multirow{3}{*}{contriever} & AGGD& \textbf{97.14} & \textbf{26.51} & - & - & 1.2\\  & Hot Flip& 96.05 & 20.9 & - & \textbf{0.18} & 0.48\\  & Random& 90.5 & 11.8 & - & 0.14 & \textbf{1.39}\\ \cline{2-7}
\hline
\multirow{3}{*}{contriever-msmarco}& AGGD& \textbf{27.76} & \textbf{83.71} & - & - & 0.28\\  & Hot Flip& 25.23 & 79.05 & - & 1.02 & 2.28\\  & Random& 26.24 & 78.38 & - & \textbf{2.05} & \textbf{7.16}\\ \cline{2-7}
\hline
\multirow{3}{*}{ance} & AGGD& 26.43 & 4.88 & \textbf{100.0} & - & - \\  & Hot Flip& \textbf{28.48} & 4.92 & 99.99 & - & - \\  & Random& 9.86 & \textbf{6.39} & 99.3 & - & - \\ \cline{2-7}
\hline
\multirow{3}{*}{dpr-multi} & AGGD& 15.78 & 3.67 & - & \textbf{87.84} & 22.47\\  & Hot Flip& \textbf{23.43} & \textbf{7.7} & - & 84.18 & \textbf{38.41}\\  & Random& 18.43 & 4.15 & - & 79.35 & 10.41\\ \cline{2-7}
\hline
\multirow{3}{*}{dpr-single} & AGGD& 10.85 & \textbf{9.99} & - & \textbf{29.16} & \textbf{89.46}\\  & Hot Flip& 11.87 & 9.45 & - & 27.91 & 87.72\\ & Random& \textbf{22.02} & 7.37 & - & 20.5 & 82.55\\ \cline{2-7}
\hline
\end{tabular}
\caption{Attack transferability across models on NQ dataset.}
\label{tab: transferability}
\end{table*}

\section{Experimental Detail}

\begin{table*}[t!]
\centering
\scriptsize
\begin{tabular}{lcc|c|c|cc}
\toprule
& \multirow{2}{*}{Domain}& \multirow{2}{*}{Dataset}& Train & Dev & \multicolumn{2}{c}{Test}\\
\cline{4-7}
&& &\# Pair & \# Query& \# Query & \# Corpus\\
\midrule
\multirow{2}{*}{In-domain}&
Web & MS MARCO \cite{nguyen2016ms}& 532,761& -& 6,980& 8,841,823\\\cline{2-7}
&\multirow{1}{*}{Wikipedia} & NQ \cite{kwiatkowski2019natural} & 132,803& -&3,452& 2,681,468\\\midrule
&\multirow{1}{*}{Bio Medical}
& NFCorpus \cite{boteva2016full} & 110,575& 324 & 323& 3,633\\ \cline{2-7}
\multirow{5}{*}{Out-of-domain}&
Quora &Quora& -&5,000&10,000&522,931\\\cline{2-7}
&\multirow{2}{*}{Scientific }& SCIDOCS \cite{cohan2020specter}& -&-&1,000&25,657\\
&&SciFact \cite{wadden2020fact}& 920& -&300& 5,183\\\cline{2-7}
&Finance& FiQA-2018 \cite{maia201818}& 14,166& 500& 648& 57,638\\
\bottomrule
\end{tabular}
\caption{Dataset Statistics. More statistics can be found in \cite{zhao2024dense, thakur2021beir}. In our experiments, we use MS MARCO and NQ datasets to train the adversarial passages and evaluate the attack on the unseen test queries from these two datasets for in-domain attack evaluation. The remaining 5 datasets (NFCorpus, Quora, SCIDOCS, SciFact, FiQA-2018) are used for out-of-domain evaluation when injecting the adversarial passages generated from the MS MARCO and NQ dataset into the corpora of these out of domain datasets.}
\label{tab:dataset statistics}
\end{table*}
\subsection{Dataset}\label{app: dataset detail}
\begin{itemize}
\item \textbf{MS MARCO} \cite{nguyen2016ms} contains a large amount of queries with annotated relevant passages from Web documents. 

\item \textbf{Natural Questions (NQ)} \cite{kwiatkowski2019natural} contains Google search queries with annotations from the top-ranked Wikipedia pages. 
\end{itemize}
These two datasets have been widely used for evaluating dense retrieval models. 

The statistics of all the datasets used in evaluation are summarized in Table \ref{tab:dataset statistics}.

\subsection{Retrievers} \label{app: retriever detail}
We experimented with the following retrievers:
\begin{itemize}
\item \textbf{Dense Passage Retriever(DPR)} \cite{karpukhin2020dense} is a two-tower bi-encoder trained with a single BM25 hard negative and in-batch negatives. It has been used as the retrieval component of many Retrieval-Augmented Generation (RAG) models \cite{lewis2020retrieval}. In our paper, we use both the open-sourced Multi model (DPR-mul), which is a bert-base-uncased model trained on four QA datasets (NQ \cite{kwiatkowski2019natural}, TriviaQA \cite{joshi2017triviaqa}, WebQuestions \cite{berant2013semantic} and CuratedTREC \cite{baudivs2015modeling})
and the single NQ model (DPR-nq). 
\item \textbf{ANCE \cite{xiong2020approximate}} is a bi-encoder that generates hard negatives using an approximate Nearest Neighbor (ANN) index of the corpus. The index is continuously updated in parallel to identify challenging negative examples for the model during fine-tuning.

\item \textbf{Contriever \cite{gautier2022unsupervised}} is an unsupervised dense retriever using contrastive learning. It leverages the BERT architecture to encode both queries and documents. Contriever-MS (Contriever fine-tuned on MS MARCO) is a version of the Contriever model that has been fine-tuned using the MS MARCO dataset, which provides large-scale, supervised training data.

\end{itemize}

\section{Knowledge Poisoning Attacks to RAG} \label{apdx: attacking RAG}
\begin{table}[h]
\centering
\scriptsize
\begin{tabular}{|c|c|c|c|c|c|c|c|}
\hline
& LLaMa-2-7B & LLaMa-2-13B & Vicuna-7B &Vicuna-13B& Vicuna-33B & GPT-3.5 & GPT-4  \\ \hline
\multicolumn{8}{|c|}{\textbf{MS MARCO}} \\ \hline
AGGD & \textbf{0.92(0.00)} & \textbf{0.92(0.00)} & \textbf{0.92(0.00)} & \textbf{0.92(0.00)} & \textbf{0.92(0.00)} & \textbf{0.92(0.00)} & \textbf{0.92(0.00)} \\ \hline
HotFlip & 0.90(0.01) & 0.90(0.01) & 0.90(0.01) & 0.90(0.01) & 0.90(0.01) & 0.90(0.01) & 0.90(0.01) \\ \hline
\multicolumn{8}{|c|}{\textbf{NQ}} \\ \hline
AGGD & 1.00(0.00) & 1.00(0.00) & 1.00(0.00) & 1.00(0.00) & 1.00(0.00) & 1.00(0.00) & 1.00(0.00) \\ \hline
hotflip & 1.00(0.00) & 1.00(0.00) & 1.00(0.00) & 1.00(0.00) & 1.00(0.00) & 1.00(0.00) & 1.00(0.00) \\ \bottomrule
\end{tabular}
\caption{PoisonedRAG performance on multiple dataset and LLMs using Hot Flip and AGGD evaluated using F1 score. }
\label{table: PoisonedRAG F1 result}
\end{table}

\paragraph{Retrieval-Augmented Generation} \cite{karpukhin2020dense, lewis2020retrieval, borgeaud2022improving, thoppilan2022lamda} augments LLMs with external knowledge retrieved from a knowledge base to improve their ability to generate accurate and up-to-date content. There are three components in RAG: knowledge base, the retriever and the LLM. The knowledge base contains a large corpus collected from various domains such as Wikipedia \cite{thakur2021beir}, Fiance \cite{loukas2023making} and Biomedical articles \cite{roberts2020trec}. Given a user question, the retriever uses a text encoder to compute the embedding vector. A set of $k$ retrieved texts from the knowledge base with the highest similarity to the question are then retrieved, which can be used by the LLM to generate content. 

RAG enables LLMs to incorporate more current knowledge by regularly updating the knowledge base. However, this also introduces potential security concerns: maliciously crafted content could be injected into the database during updates, which might then be retrieved by the LLM to generate false, biased or harmful output.

\paragraph{Threat Model}
We consider a scenario where the attacker does not have direct access to the database, but can access the weights of the retrieval model. This scenario is realistic, as the database is likely hosted on a secure system, while the retriever used may be an open-access LLM. We assume the attacker can inject a few carefully crafted corpus into the knowledge base. 
\paragraph{Metric}
We use the same metrics, Attack Success Rate (ASR) and F1, as used in \cite{zou2024poisonedrag}. However, since  ASR is task-specific, it might not be exactly the same as defined for corpus poisoning attacks. In PoisonedRAG, the definitions are given as follows.
\begin{itemize}
    \item \textbf{ASR} The fraction of target questions whose answers are the attacker-chosen target answers. 
\end{itemize}
Additionally, we report average F1-Score over different target
questions in Table \ref{table: PoisonedRAG F1 result}, which measures the trade-off between Precision and Recall. Specifically, 
\begin{equation}
\text{F1 score} = \frac{2\cdot \text{Precision}\cdot\text{Recall}}{\text{Precision}+\text{Recall}}
\end{equation}
where Precision is defined as the fraction of poisoned
texts among the top-$k$ retrieved ones for the target question and  Recall is defined as the fraction of poisoned texts among the $N$
poisoned ones that are retrieved for the target question.
 A higher F1-Score means more poisoned texts are retrieved.
 \paragraph{Hyperparameters}
We inject 1 adversarial poisoned text for each target question and test on all 100 test data. The results are averaged over 4 random runs. 
\section{Examples of the output passage from AGGD}

\begin{table*}[h]
\centering
\begin{tabular}{p{8cm}|p{8cm}}
\toprule
Examples from attacking Contriever & Examples from attacking contriever-MS \\
\midrule
\ttfamily{["\#\#nius", "half", "melting", "definite", "[MASK]", "favor", "gma", "who", "should", "jew", "rain", "\#\#pi", "\#\#ntial", "upper", "kevin", "perrin", "\#\#gren", "blew", "demonstrators", "remnants", "shelters", "that", "\#\#ntial", "hue", "lest", "rainfall", "where", "rains", "rain", "heavy"]} & \ttfamily{["hostage", "tightly", "handful", "herbs", "where", "packets", "were", "\#\#ssen", "dealers", "overnight", "symbol", "\#\#atic", "\#\#ised", "adventure", "bail", "\#\#alis", "\#\#kari", "[CLS]", "convicted", "ga", "\#\#oit", "peace", "restore", "lifespan", "discrimination", "(", "maha", "\#\#ter", "bank", "\#\#ees"]}\\
\bottomrule
\end{tabular}
\caption{Examples of adversarial passages}
\label{tab:examples_of_adversarial_passages}
\vspace{-0.5cm}
\end{table*}
\end{document}